\documentclass[aps,showpacs,amsmath,amssymb,pre,superscriptaddress,twocolumn]{revtex4}
\usepackage{graphicx}
\usepackage{epsfig}
\usepackage{ulem}
\usepackage{amsmath}
\usepackage{amssymb}
\usepackage{color}

\begin{document}

\title{A systematic study of electron or hole transfer along DNA dimers, trimers and polymers}
%\thanks{A footnote to the article title}

\author{Constantinos Simserides}
\email{csimseri@phys.uoa.gr}
\homepage{http://users.uoa.gr/~csimseri/}
\affiliation{National and Kapodistrian University of Athens, Faculty of Physics,
Panepistimiopolis, 15784 Zografos, Athens, Greece}
\date{\today}% It is always \today, today,


% 82.       Physical chemistry and chemical physics
% 82.39.-k 	Chemical kinetics in biological systems
% 82.39.Jn 	Charge (electron, proton) transfer in biological systems
% 82.39.Pj 	Nucleic acids, DNA and RNA bases

% 87. 	    Biological and medical physics
% 87.14.-g 	Biomolecules: types
% 87.14.gk 	DNA
% 87.15.-v 	Biomolecules: structure and physical properties
% 87.15.A- 	Theory, modeling, and computer simulation

% 72. 	    Electronic transport in condensed matter
% 72.80.-r 	Conductivity of specific materials
% 72.80.Le 	Polymers; organic compounds (including organic semiconductors)

% 73. 	    Electronic structure and electrical properties of surfaces, interfaces, thin films, and low-dimensional structures
% 73.63.-b 	Electronic transport in nanoscale materials and structures

\begin{abstract}
A systematic study of electron or hole transfer along DNA dimers, trimers and polymers is presented with a tight-binding approach at the base-pair level, using the relevant on-site energies of the base-pairs and the hopping parameters between successive base-pairs.
A system of $N$ coupled differential equations is solved numerically with the eigenvalue method, allowing the temporal and spatial evolution of electrons or holes along a $N$ base-pair DNA segment to be determined.
Useful physical quantities are defined and calculated including the maximum transfer percentage $p$ and the \textit{pure} maximum transfer rate $\frac{p}{T}$ for cases where a period $T$ can be defined, as well as the \textit{pure} mean carrier transfer rate $k$ and the speed of charge transfer $ u = k d$, where $d = N \times$ 3.4 {\AA} is the charge transfer distance. The inverse decay length $\beta$ used for the exponential fit $k = k_0 \textrm{exp}(-\beta d)$ and the exponent $\eta$ used for the power law fit $k = k_0' N^{-\eta}$ are computed.
The electron and hole transfer along polymers including poly(dG)-poly(dC), poly(dA)-poly(dT), GCGCGC..., ATATAT... is studied, too.
$\beta$ falls in the range $\approx$ 0.2 - 2 {\AA}$^{-1}$, $k_0$ is usually 10$^{-2}$-10$^{-1}$ PHz although, generally, it falls in the wider range 10$^{-4}$-10 PHz.
$\eta$ falls in the range $\approx$ 1.7 - 17, $k_0'$ is usually $\approx 10^{-2}$-10$^{-1}$ PHz, although generally, it falls in the wider range $\approx 10^{-4}$-10$^3$ PHz.
Finally, the results are compared with the predictions of Wang~\textit{et al.} Phys. Rev. Lett. \textbf{93}, (2004) 016401,
as well as experiments, including
Murphy~\textit{et al.} Science \textbf{262}, 1025 (1993);
Arkin~\textit{et al.} Science \textbf{273}, 475 (1996);
Giese~\textit{et al.} Angew. Chem. Int. Ed. \textbf{38}, 996 (1999);
Giese~\textit{et al.} Nature \textbf{412}, 318 (2001).
This method allows to assess the extent at which a specific DNA segment can serve as an efficient medium for charge transfer.
\end{abstract}

\maketitle

%%%%%%%%%%%%%%%%%%%%%%%%%%%%%%%%%%%%%%%%%%%%%%%%%%%%%%%%%%%%%%%%%%%%
\section{\label{sec:introduction} Introduction}
%%%%%%%%%%%%%%%%%%%%%%%%%%%%%%%%%%%%%%%%%%%%%%%%%%%%%%%%%%%%%%%%%%%%
It is not the purpose of this work to discuss the reasons charge transfer along deoxyribonucleic acid (DNA) is crucial
for molecular biology, genetics, and nanotechnology. The interested reader may consult valuable reviews e.g. Refs.
\cite{GenereuxBarton:2010, Giese:2002, Endres:2004}.
The aim is to present a convenient way to quantify electron or hole transport along specific DNA segments using a simple tight-binding approach
which can easily be implemented by interested colleagues.

In this spirit, much interest has been devoted recently to the study of the tight-binding parameters which are relevant to charge transport along DNA~\cite{Endres:2004,HKS:2010-2011,Senthilkumar:2005,YanZhang:2002,SugiyamaSaito:1996,HutterClark:1996,ZhangLiEtAl:2002,
LiCaiSevilla:2001,LiCaiSevilla:2002,ShuklaLeszczynski:2002,Varsano:2006,Voityuk:2001,Migliore:2009,Kubar:2008,Ivanova:2008}.
Among these works, parameters either for electrons or holes were calculated~\cite{HKS:2010-2011},
employing a novel parametrization within a simplified linear combination of atomic orbitals (LCAO) method \cite{HKS:2009}.
Specifically, the $\pi$ molecular structure of the four DNA bases [adenine (A), thymine (T), guanine (G), cytosine (C)] was investigated,
the HOMO (highest occupied molecular orbital) and the LUMO (lowest unoccupied molecular orbital) wave functions and energies calculated, and
then used to obtain the wave functions of the two B-DNA base-pairs (A-T and G-C)
using a linear combination of molecular orbitals (LCMO) of the complementary bases (A and T for A-T, G and C for G-C).
Then the complete set of the charge transfer parameters
(between successive base-pairs and also between neighboring bases considering all possible combinations)
both for electrons and holes was estimated.
Conclusively, to date all these parameters either for electrons (traveling through LUMOs) or for holes (traveling through HOMOs) are available in the literature. In the present article we are going to use them to study the
temporal and spatial evolution of an electron or a hole as it is transferred along DNA.

The transfer of electrons or holes along DNA can be described at either
(I) the base-pair level or
(II) the single base level~\cite{HKS:2010-2011}.
One has to know the relevant on-site energies of either
(I) the base-pairs or
(II) the single bases.
In addition, one has to provide, the hopping parameters between either
(I) successive base-pairs or
(II) neighboring bases taking all possible combinations into account
[(IIa) successive bases in the same strand,
(IIb) complementary bases within a base-pair,
(IIc) diagonally located bases of successive base-pairs in opposite strands].
Knowing these parameters,
to calculate the temporal and spatial evolution of carriers along a $N$ base-pair segment of DNA
one has to solve a system of either
(I) $N$ or
(II) $2N$
coupled differential equations.
In the present work we use the simplest approach (I) which is analyzed in Section~\ref{sec:charge-transfer}.
Next, in Section~\ref{sec:resdis} we solve numerically the relevant system of coupled differential equations and examine charge transfer in DNA dimers, trimers and polymers. We also compare our results with available experiments.
Finally, in Section~\ref{sec:conclusion} we state our conclusions.

\section{Charge transfer in B-DNA. Description at the base-pair level}%%%%
\label{sec:charge-transfer}%%%%%%%%%%%%%%%%%%%%%%%%%
We assume that extra electrons inserted in DNA travel through LUMOs, while inserted holes travel through HOMOs.
We symbolize successive base-pairs of double-stranded DNA by $\ldots, \mu-1, \mu, \mu+1, \ldots$ and
the description can be done either for HOMOs or for LUMOs \cite{HKS:2010-2011}.
For a description at a base-pair level we need the HOMO or the LUMO on-site energies of the base-pairs
and the corresponding hopping parameters between successive base-pairs.
The relevant parameters are given in Table~\ref{table:bpHL} and Table~\ref{table:interbpTP}.

\begin{widetext}

\begin{table}[h!]
\caption{The on-site energies $E^{bp}_{H/L}$ for the two possible base-pairs A-T and G-C, calculated by various authors.
%{\color{red}${E^{bp \; \textrm{used}}_{H/L}}\mathrm{used}$}
${E^{bp \; \textrm{used}}_{H/L}}$
are the values actually used for the solution of Eq.~\ref{TBbp} in this article.
The first $\pi$-$\pi^*$ transition energies $E_{\pi-\pi^*}$ for the two B-DNA base-pairs are also shown.
Except for Ref.~\cite{HKS:2010-2011} these are \textit{ab initio} calculations which tend to overestimate the first $\pi$-$\pi^*$ transition energy.
All energies are given in eV.}
\centerline{
\begin{tabular}{|l|c|c|c|} \hline
B-DNA base-pair  & A-T     &   G-C   & reference             \\  \hline  \hline
$E_H^{bp}$       & $-$8.3  & $-$8.0  & \cite{HKS:2010-2011}  \\  \hline
$E_L^{bp}$       & $-$4.9  & $-$4.5  & \cite{HKS:2010-2011}  \\  \hline
$E_{\pi-\pi^*}$  &     3.4 &    3.5  & \cite{HKS:2010-2011}  \\  \hline
$E_H^{bp \; \mathrm{first \; pr.}}$     & $-$(7.8-8.2) & $-$(6.3-7.7) & \cite{SugiyamaSaito:1996,HutterClark:1996,ZhangLiEtAl:2002,LiCaiSevilla:2001,LiCaiSevilla:2002,ShuklaLeszczynski:2002} \\ \hline
$E_{\pi-\pi^*}^{\mathrm{first \; pr.}}$ &         6.4  &     4.3-6.3  & \cite{ShuklaLeszczynski:2002,Varsano:2006} \\ \hline
${E^{bp \; \textrm{used}}_{H}}$&   8.3 &    8.0 & \cite{HKS:2010-2011}  \\  \hline
${E^{bp \; \textrm{used}}_{L}}$&$-$4.9 & $-$4.5 & \cite{HKS:2010-2011}  \\  \hline
\end{tabular}  } \label{table:bpHL}
\end{table}

\begin{table}[h!]
\caption{The hopping parameters between successive base-pairs for all possible combinations. $t_H^{bp}$ ($t_L^{bp}$) refers to hole (electron) hopping through HOMOs (LUMOs). The notation is given in the text. The values listed
in Table 3 of Ref.~\cite{HKS:2010-2011},
in Table II or Ref.~\cite{Voityuk:2001},
in Table 5 (``Best Estimates'') of Ref.~\cite{Migliore:2009},
in Table 4 of Ref.~\cite{Kubar:2008} (two estimations given),
in Table 2 of Ref.~\cite{Ivanova:2008}, and
the values extracted approximately from Fig.~4 of Ref.~\cite{Endres:2004} are shown.
These quantities represent the parameters
$t^{bp (\mu;\mu\pm1)}_{H/L}$ which appear in Eq.~(\ref{TBbp}).
Finally, ${t^{bp \; \textrm{used}}_{H/L}}$ are the parameters actually used in this work
for the solution of Eq.~\ref{TBbp}. All hopping integrals $t_{H/L}^{bp}$ are given in meV.}
\centerline{
\begin{tabular}{|c|r|r|r|r|r|r|r||r|r|r|} \hline
Base-pair& $t_H^{bp}$ & $|t_H^{bp}|$ & $t_H^{bp}$ & $t_H^{bp}$ & $t_H^{bp}$ & $t_H^{bp}$ & $t_H^{bp \; \textrm{used}}$ & $t_L^{bp}$ & $t_L^{bp}$ & $t_L^{bp \; \textrm{used}}$ \\
sequence & \cite{HKS:2010-2011} & \cite{Voityuk:2001} & \cite{Endres:2004} & \cite{Migliore:2009} & \cite{Kubar:2008} & \cite{Ivanova:2008} & & \cite{HKS:2010-2011} & \cite{Endres:2004} &  \\ \hline \hline
AA, TT& $-$8& 26& $-$25 &8-17& 19(19)&22&    20  &$-$29& 35 & $-$29 \\ \hline
    AT&   20& 55&       &    & 47(74)&37& $-$35  &  0.5&    &     0.5  \\ \hline
AG, CT& $-$5& 25& $-$50 &    & 35(51)&43&    30  &    3& 35 &     3 \\ \hline
AC, GT&    2& 26&       &    & 25(38)&20& $-$10  &   32&    &    32 \\ \hline
    TA&   47& 50&       &    & 32(68)&52& $-$50  &    2&    &     2 \\ \hline
TG, CA& $-$4& 27&       &    & 11(11)&25&    10  &   17&    &    17 \\ \hline
TC, GA&$-$79&122&$-$160 &    &71(108)&60&   110  & $-$1& 35 &  $-$1 \\ \hline
GG, CC&$-$62& 93&$-$140 &  75&72(101)&63&   100  &   20& 35 &    20 \\ \hline
    GC&    1& 22&       &    & 20(32)&22& $-$10  &$-$10&    & $-$10 \\ \hline
    CG&$-$44& 78&       &    & 51(84)&74&    50  & $-$8&    &  $-$8 \\ \hline
\end{tabular} }   \label{table:interbpTP}
\end{table}

\end{widetext}

We use the notation YX to denote two successive base-pairs,
according to the following convention for the DNA strands orientation
\begin{eqnarray}
   &\vdots& \nonumber \\
5' &      & 3' \nonumber \\
\textrm{Y}  &   -  & \textrm{Y}_{\textrm{compl}} \nonumber \\
\textrm{X}  &   -  & \textrm{X}_{\textrm{compl}} \nonumber  \\
3' &      & 5' \nonumber \\
   &\vdots&
\label{bpdimer}
\end{eqnarray}
X, X$_{\textrm{compl}}$, Y, Y$_{\textrm{compl}}$ denote DNA bases.
X$_{\textrm{compl}}$ (Y$_{\textrm{compl}}$) is the complementary base of X (Y).
In other words, the notation YX means that the bases Y and X of
two successive base-pairs are located at the same strand in the direction $5'-3'$.
X-X$_{\textrm{compl}}$ is the one base-pair and
Y-Y$_{\textrm{compl}}$ is the other base-pair,
separated and twisted by 3.4 {\AA} and $36^{\circ}$, respectively,
relatively to the first base-pair.
For example, the notation AC denotes that the base-pair dimer consists of an
adenine-thymine and a cytosine-guanine base-pair, where
one strand contains A and C in the direction $5'-3'$ and the complementary
strand contains T and G in the direction $3'-5'$.

For a description at the base-pair level,
the time-dependent single carrier (hole/electron)
wave function of the DNA segment of interest,
$\Psi^{DNA}_{H/L}({\bf r},t)$,
is considered as a linear combination of base-pair wave functions with
time-dependent coefficients, i.e.,
\begin{equation}\label{psi-total}
\Psi^{DNA}_{H/L}({\bf r},t) = \sum_{\mu=1}^{N} A_{\mu}(t) \;
\Psi^{bp(\mu)}_{H/L}({\bf r}),
\end{equation}
where $\Psi^{bp(\mu)}_{H/L}({\bf r})$ is the $\mu^{th}$
base-pair's HOMO or LUMO wave function ($H/L$) and
the sum is extended over all base-pairs of the DNA segment under consideration.
Hence, $|A_{\mu}(t)|^2$ gives the probability of finding the carrier at the base-pair $\mu$
at the time $t$.

Starting from the time-dependent Schr\"{o}dinger equation,
\begin{equation}\label{tdse}
i \hbar \frac{\partial \Psi^{DNA}_{H/L}({\bf r},t)}{\partial t} =H^{DNA}\Psi^{DNA}_{H/L}({\bf r},t),
\end{equation}
using Eq.~\ref{psi-total}, and following the procedure and the assumptions described in detail by Hawke {\it et al.}~\cite{HKS:2010-2011},
one obtains that the time evolution of the coefficients $A_\mu(t)$
obeys the tight-binding system of differential equations
\begin{equation}\label{TBbp}
i \hbar \frac{ dA_{\mu}} {dt} = E^{bp (\mu)}_{H/L} A_{\mu}
+ t^{bp (\mu;\mu-1)}_{H/L} A_{\mu-1}
+ t^{bp (\mu;\mu+1)}_{H/L} A_{\mu+1}.
\end{equation}
$E^{bp (\mu)}_{H/L}$ is the HOMO/LUMO on-site energy of base-pair $\mu$, and
$t^{bp (\mu; \mu ')}_{H/L}$ is the hopping parameter between base-pair $\mu$ and base-pair $\mu '$.

Hence, one can use
$E^{bp (\mu)}_{H/L}$ (cf. Table~\ref{table:bpHL}) and
$t^{bp(\mu;\mu ')}_{H/L}$ (c.f. Table~\ref{table:interbpTP})
computed already by many authors in order to numerically solve the system of equations (\ref{TBbp}) and obtain,
through $A_{\mu}(t)$, the time evolution of a carrier propagating along the DNA segment of interest.
Regarding the tight-binding description of hole transport,
the corresponding tight-binding parameters should be taken with the opposite sign of
the calculated on-site energies and transfer hopping integrals~\cite{Senthilkumar:2005}.
This means that for describing hole transport at the base-pair level,
the on-site energies $E_H^{bp}$ presented in the second row of Table~\ref{table:bpHL} and
the hopping transfer integrals $t_H^{bp}$ presented in the second column of Table~\ref{table:interbpTP}
should be used with opposite signs to provide the tight-binding parameters of Eq.~\ref{TBbp}.

The on-site energies $E^{bp}_{H/L}$ for the two possible base-pairs A-T and G-C, calculated by various authors, are listed in Table~\ref{table:bpHL}. ${E^{bp \; \textrm{used}}_{H/L}}$ are the values actually used for the solution of Eq.~\ref{TBbp} in this article.
The hopping parameters $t_{H/L}^{bp}$ for all possible combinations of successive base-pairs, calculated by various authors, are given in Table~\ref{table:interbpTP}.
${t^{bp \; \textrm{used}}_{H/L}}$ are the values actually used for the solution of Eq.~\ref{TBbp} in this article.
Due to the symmetry between base-pair dimers YX and X$_{\textrm{compl}}$Y$_{\textrm{compl}}$,
the number of different hopping parameters is reduced from sixteen to ten.
In Table \ref{table:interbpTP} base-pair dimers
exhibiting the same transfer parameters are listed together in the first column.
We include in Table~\ref{table:interbpTP} the values listed:
in Table 3 of Ref.~\cite{HKS:2010-2011},
in Table II or Ref.~\cite{Voityuk:2001},
in Table 5 (``Best Estimates'') of Ref.~\cite{Migliore:2009},
in Table 4 of Ref.~\cite{Kubar:2008} (two estimations given),
in Table 2 of Ref.~\cite{Ivanova:2008}, and
the values extracted approximately from Fig.~4 of Ref.~\cite{Endres:2004}.
In Refs.~\cite{Kubar:2008,Migliore:2009,Ivanova:2008} all values given are positive,
in Ref.~\cite{Voityuk:2001} the authors explicitly state that they quote absolute values,
while in Refs.~\cite{HKS:2010-2011,Endres:2004} the sign is included.
In Ref.~\cite{HKS:2010-2011} all $t_H^{bp}$ and $t_L^{bp}$ have been calculated, while
in Ref.~\cite{Endres:2004} only the values of $t_H^{bp}$ for a few cases are approximately given.
According to Ref.~\cite{BlancafortVoityuk:2006} the approximation used in Ref.~\cite{Voityuk:2001} in general overestimates the transfer integrals. Summarizing, taking all the above into account, we use the values
${E^{bp \; \textrm{used}}_{H/L}}$ and ${t^{bp \; \textrm{used}}_{H/L}}$.

To solve Eq.~~\ref{TBbp} we define the vector matrix
\begin{equation}\label{x}
\vec{x}(t) = \left[
\begin{array}{c}
A_1(t) \\
A_2(t) \\
\vdots \\
A_N(t)  \end{array} \right]
\end{equation}
and therefore Eq.~~\ref{TBbp} reads
\begin{equation}\label{xdotmathcalAx}
\dot{\vec{x}}(t) = \widetilde{\mathcal{A}} \vec{x}(t),
\end{equation}
\begin{equation}\label{mathcalA}
\widetilde{\mathcal{A}} = - \frac{i}{\hbar} \textrm{A},
\end{equation}

\begin{widetext}
\begin{equation}\label{A}
%\scriptsize{
\textrm{A} = \left[
\begin{array}{ccccccc}
  E^{bp(1)}_{H/L}&t^{bp(1;2)}_{H/L}& 0                &\cdots&     0 &      0 & 0    \\
t^{bp(2;1)}_{H/L}&  E^{bp(2)}_{H/L}& t^{bp(2;3)}_{H/L}&\cdots&     0 &      0 & 0    \\
\vdots           & \vdots          & \vdots           &\vdots&\vdots & \vdots &\vdots\\
0      & 0     & 0      &\cdots&t^{bp(N-1;N-2)}_{H/L}&  E^{bp(N-1)}_{H/L} & t^{bp(N-1;N)}_{H/L}    \\
0      & 0     &  0     &\cdots&                   0 &t^{bp(N;N-1)}_{H/L} & E^{bp(N)}_{H/L} \end{array} \right].
%}
\end{equation}

\end{widetext}

The matrix $\textrm{A}$ is a symmetric tridiagonal matrix. We solve Eq.~\ref{xdotmathcalAx} using the {\it eigenvalue method}, i.e. we look for solutions of the form $\vec{x}(t) = \vec{v} e^{\tilde{\lambda} t} \Rightarrow \dot{\vec{x}}(t) = \tilde{\lambda} \vec{v} e^{\tilde{\lambda} t}$. Hence, Eq.~\ref{xdotmathcalAx} reads
\begin{equation}\label{mathcalA}
\widetilde{\mathcal{A}} \vec{v} = \tilde{\lambda} \vec{v},
\end{equation}
or
\begin{equation}\label{textrmA}
\textrm{A} \vec{v} = \lambda \vec{v},
\end{equation}
with $ \tilde{\lambda} = - \frac{i}{\hbar} \lambda $. In other words, we solve numerically an eigenvalue problem.
We also compare our numerical results with known analytical solutions in some simple known cases.

For example, in the cases of poly(dA)-poly(dT) or poly(dG)-poly(dC) DNA segments the matrix $\textrm{A}$ becomes a symmetric tridiagonal uniform matrix
\begin{equation}\label{Auniform}
\textrm{A} = \left[
\begin{array}{ccccccc}
E^{bp} & t^{bp} & 0 &\cdots& 0 & 0 & 0 \\
t^{bp} & E^{bp} & t^{bp} &\cdots& 0 & 0 & 0 \\
\vdots&\vdots&\vdots&\vdots&\vdots&\vdots&\vdots\\
0 & 0 & 0 &\cdots& t^{bp} & E^{bp} & t^{bp} \\
0 & 0 & 0 &\cdots& 0 & t^{bp} & E^{bp} \end{array} \right]
\end{equation}
whose eigenvalues are $\lambda_k = E^{bp} + 2 t^{bp} \textrm{cos}(\frac{k\pi}{N+1})$, where $k=1,2,\dots,N$. Hence,
all eigenvalues are real and distinct being the matrix symmetric ($\textrm{A} = \textrm{A}^\textrm{T}$),
all eigenvalues are symmetric around $E^{bp}$,
for odd $N$ the trivial eigenvalue ($ = E^{bp}$) exists, and
all eigenvalues lie in the interval $(E^{bp}-2t^{bp},E^{bp}+2t^{bp})$. The corresponding eigenvectors
i.e. the $k'$ component of the $k$ eigenvector are
$u_{k' k} \propto \textrm{sin}(k' k \frac{\pi}{N+1})$,
where $k=1,2,\dots,N$ and $k'=1,2,\dots,N$.
One needs to make the substitution $k \to N+1-k$ to obtain the ascending order.

For long enough segments of DNA one could envision the cyclic case with
$ \textrm{A}(1,N) = t^{bp(1;N)}_{H/L} = \textrm{A}(N,1) = t^{bp(N;1)}_{H/L} \neq 0 $. Hence,
in the cases of {\it cyclic} poly(dA)-poly(dT) or poly(dG)-poly(dC) DNA segments the matrix $\textrm{A}$ becomes a symmetric tridiagonal uniform matrix with two ``perturbed corners'' i.e.
\begin{equation}\label{Auniform}
\textrm{A} = \left[
\begin{array}{ccccccc}
E^{bp} & t^{bp} & 0 &\cdots& 0 & 0 & t^{bp} \\
t^{bp} & E^{bp} & t^{bp} &\cdots& 0 & 0 & 0 \\
\vdots&\vdots&\vdots&\vdots&\vdots&\vdots&\vdots\\
0 & 0 & 0 &\cdots& t^{bp} & E^{bp} & t^{bp} \\
t^{bp} & 0 & 0 &\cdots& 0 & t^{bp} & E^{bp} \end{array} \right]
\end{equation}
whose eigenvalues are \cite{YuehCheng:2008}
$\lambda_k = E^{bp} + 2t^{bp}\textrm{cos}\theta_k$, $\theta_k = \frac{2 k \pi}{N}$, $k = 1, 2, \dots, N$.

Having checked that the normalized eigenvectors $\vec{v_k}$ corresponding to the eigenvalues $\lambda_k$ of Eq.~\ref{textrmA} are linearly independent, the solution of our problem is
\begin{equation}
\vec{x}(t) = \sum_{k=1}^{N} c_k \vec{v_k} e^{-\frac{i}{\hbar} \lambda_k t}.
\end{equation}

The initial conditions used usually in this article are
\begin{equation}\label{x0}
\vec{x}(0) = \left[
\begin{array}{c}
A_1(0) \\
A_2(0) \\
\vdots \\
A_N(0)  \end{array} \right]
=
\left[
\begin{array}{c}
1 \\
0 \\
\vdots \\
0  \end{array} \right]
\end{equation}
which means that we initially place the carrier in base-pair 1 and we want to see how the carrier will evolve, time passing. From the initial conditions we determine $c_i(t)$. In some cases, however, comparing with a particular experiment, we need to put the carrier initially at the base-pair indicated by the authors of the experimental work.

%%%%%%%%%%%%%%%%%%%%%%%%%%%%%%%%%%%%%%%%%%%%%%%%%%%%%%%%%%%%%%%%%%%%%%%%%%%%%%%%%%%%%%%%%%%%%
\section{Dimers, trimers and polymers} %%%%%%%%%%%%%%%%%%%%%%%%%%%%%%%%%%%%%%%%%%%%%%%%%%%%%%
\label{sec:resdis}%%%%%%%%%%%%%%%%%%%%%%%%%%%%%%%%%%%%%%%%%%%%%%%%%%%%%%%%%%%%%%%%%%%%%%%%%%%
%%%%%%%%%%%%%%%%%%%%%%%%%%%%%%%%%%%%%%%%%%%%%%%%%%%%%%%%%%%%%%%%%%%%%%%%%%%%%%%%%%%%%%%%%%%%%
We emphasize that our calculations below refer to \textit{pure} carrier transfer rates extracted from the probabilities to find the carrier at a particular monomer of interest after having placed it initially (for time zero) at another monomer, i.e. refer directly to the solution of Eq.~(\ref{TBbp}).
%These are extracted from formulas of the type given in Eq.~(\ref{meantransferrateN}) or the relevant for dimers and trimers Eq.~(\ref{meantransferrate2}) and Eq.~(\ref{meantransferrate3}) below.
We do not take into account the influence of other factors such as the density of states, the environment etc.

%%%%%%%%%%%%%%%%%%%%%%%%%%
\subsection{Dimers}%%%%%%%
\label{subsec:dimers}%%%%%%
%%%%%%%%%%%%%%%%%%%%%%%%%%
For any dimer, the solution of our problem is
\begin{equation}\label{dimersolution}
\vec{x}(t) = \sum_{k=1}^{2} c_k \vec{v_k} e^{-\frac{i}{\hbar} \lambda_k t}.
\end{equation}
Let us suppose that $\lambda_2 \ge \lambda_1$.
We are interested in the quantities $|A_{\mu}(t)|^2, \mu = 1, 2$ (cf. Eq.~\ref{psi-total} and Eq.~\ref{x}) since they provide the probabilities of finding the carrier at the base-pair $\mu$. From Eq.~\ref{dimersolution}, we obtain the period of $|A_{\mu}(t)|^2, \mu = 1, 2$ \begin{equation}\label{period}
T = \frac{h}{\lambda_2 - \lambda_1}.
\end{equation}

\subsubsection{A dimer consisting of identical monomers}
\label{dimersidmon}
Maybe in the simplest case one could imagine, suppose that we have a dimer consisting of two identical monomers
with purine on purine and pyrimidine on pyrimidine, i.e. GG $\equiv$ CC or AA $\equiv$ TT e.g.
\begin{eqnarray}
5' &      & 3' \nonumber \\
\textrm{G}  &   -  & \textrm{C} \nonumber \\
\textrm{G}  &   -  & \textrm{C} \nonumber  \\
3' &      & 5'
\label{GG}
\end{eqnarray}
so that
\begin{equation}\label{Auniform}
\textrm{A} = \left[
\begin{array}{cc}
E^{bp} & t^{bp}  \\
t^{bp} & E^{bp}  \end{array} \right]
\end{equation}
with eigenvalues
\begin{equation}\label{EigenvIdMonomers}
\lambda_{1,2} = E^{bp} \mp t^{bp}
\end{equation}
and corresponding normalized eigenvectors
\begin{equation}
\vec{v_1} = \left[
\begin{array}{c}
-\sqrt{2}/2 \\
\sqrt{2}/2 \end{array} \right]
,
\vec{v_2} = \left[
\begin{array}{c}
\sqrt{2}/2 \\
\sqrt{2}/2 \end{array} \right].
\end{equation}
Then, for initial condition
\begin{equation}
\vec{x}(0) =
\left[
\begin{array}{c}
1 \\
0 \end{array} \right],
\end{equation}
\begin{equation}\label{vecxt2}
\vec{x}(t) \!\!=\!\!
\left[
\begin{array}{c}
A_1(t) \\
A_2(t) \end{array} \right]
\!\!=\!\!
\left[
\begin{array}{r}
 \frac{1}{2} e^{-\frac{i}{\hbar} \lambda_1 t} + \frac{1}{2} e^{-\frac{i}{\hbar} \lambda_2 t}    \\
-\frac{1}{2} e^{-\frac{i}{\hbar} \lambda_1 t} + \frac{1}{2} e^{-\frac{i}{\hbar} \lambda_2 t} \end{array} \right].
\end{equation}
%To the eigenvalues $\lambda_i$ ($i = 1, 2$) of Eq.~\ref{vecxt2} correspond the frequencies $f_i$ and the periods $T_i$ defined by  $\frac{\lambda_i}{\hbar} = \frac{2\pi}{T_i} = 2 \pi f_i$, i.e.
%\begin{equation}
%\frac{f_2}{f_1} = \frac{T_1}{T_2} = \frac{\lambda_2}{\lambda_1} = \frac{E^{bp} + t^{bp}}{E^{bp} - t^{bp}}.
%\end{equation}
%In reality, as mentioned above,
We are interested in the quantities $|A_{\mu}(t)|^2, \mu = 1, 2$ since they provide the probabilities of finding the carrier at the base-pair $\mu$. From Eq.~\ref{vecxt2} we obtain
\begin{equation}\label{vecxt2squared}
\left[
\begin{array}{c}
|A_1(t)|^2 \\
|A_2(t)|^2 \end{array} \right]
=
\left[
\begin{array}{r}
\frac{1}{2} + \frac{1}{2}\textrm{cos}[\frac{(\lambda_2 - \lambda_1)t}{\hbar}] \\
\frac{1}{2} - \frac{1}{2}\textrm{cos}[\frac{(\lambda_2 - \lambda_1)t}{\hbar}] \end{array} \right].
\end{equation}

Next, suppose that we have a dimer consisting of two identical monomers with purine on pyrimidine and vice versa, i.e.,
GC or CG or AT or TA, e.g.
\begin{eqnarray}
5' &      & 3' \nonumber \\
\textrm{G}  &   -  & \textrm{C} \nonumber \\
\textrm{C}  &   -  & \textrm{G} \nonumber  \\
3' &      & 5'
\label{GC}
\end{eqnarray}
In these cases the matrix $\textrm{A}$ is given again by Eq.~\ref{Auniform} i.e. the problem is identical to the previous one.

\subsubsection{A dimer consisting of different monomers}
\label{dimersdimon}
Suppose now that we have a dimer made up of two different monomers i.e. AG $\equiv$ CT, AC $\equiv$ GT, TG $\equiv$ CA,  TC $\equiv$ GA, e.g.
\begin{eqnarray}
5' &      & 3' \nonumber \\
\textrm{G}  &   -  & \textrm{C} \nonumber \\
\textrm{A}  &   -  & \textrm{T} \nonumber  \\
3' &      & 5'.
\label{GG}
\end{eqnarray}
Then,
\begin{equation}\label{GAorTC}
\textrm{A} = \left[
\begin{array}{cc}
E^{bp1} & t^{bp}  \\
t^{bp} & E^{bp2}  \end{array} \right]
\end{equation}
with eigenvalues
\begin{equation}\label{EigenvDiMonomers}
\lambda_{1,2} = \frac{E^{bp1} + E^{bp2}}{2} \mp \sqrt{\frac{(E^{bp1} - E^{bp2})^2}{4}+{t^{bp}}^2}.
\end{equation}
Eq.~\ref{EigenvIdMonomers} is a special case of Eq.~\ref{EigenvDiMonomers} for $E^{bp1} = E^{bp2}$.

\subsubsection{Period, maximum transfer percentage, pure maximum transfer rate}
If the dimer is made up of identical monomers (different monomers), the eigenvalues are given by Eq.~\ref{EigenvIdMonomers} (Eq.~\ref{EigenvDiMonomers}).
%We depict this schematically in Fig.~\ref{dimers}.
Let us define $\Delta^{bp} = |E^{bp1} - E^{bp2}|$.
In the special case $\lambda_2 = \lambda_1 \Leftrightarrow [t^{bp} = 0 \textrm{ and } \Delta^{bp} = 0]$
(cf. Eq.~\ref{EigenvIdMonomers} and Eq.~\ref{EigenvDiMonomers}), $T\to \infty$,
i.e. the functions are constant.
Using Eq.~\ref{EigenvIdMonomers} for identical monomers
\begin{equation}\label{periodIdMonomers}
T = \frac{h}{2 |t^{bp}|}.
\end{equation}
Using Eq.~\ref{EigenvDiMonomers} for different monomers
\begin{equation}\label{perioddimers}
T = \frac{h}{\sqrt{(2t^{bp})^2 + (\Delta^{bp})^2}}.
\end{equation}
The former Eq.~\ref{periodIdMonomers} is a special case of the latter Eq.~\ref{perioddimers} for $E^{bp1} = E^{bp2}$.
Hence, in the case of identical monomers the period depends only on the hopping parameter between the base-pairs, but in the case of different monomers the period depends {\it additionally} on the energy gap between the on-site energies of the carrier in the different monomers.

From Eq.~\ref{dimersolution} we obtain the maximum transfer percentage of the carrier from base-pair 1 to base-pair 2,
$p = 4 c_1 v_{11} c_2 v_{12}$. This refers to the maximum of $|A_{2}(t)|^2$.
$v_{ij}$ is the $i$-th component of eigenvector $j$.
The maximum transfer percentage reads
\begin{equation}\label{pdimers}
p = \frac{(2t^{bp})^2}{(2t^{bp})^2+(\Delta^{bp})^2}.
\end{equation}
Hence, for identical monomers, it follows that $p = 1$, while for different monomers $p < 1$.

The \textit{pure} maximum transfer rate
%({\bf maybe not the best name}),
can be defined as
\begin{equation}\label{pperTdimers}
\frac{p}{T} = \frac{(2t^{bp})^2}{h\sqrt{(2t^{bp})^2+(\Delta^{bp})^2}}.
\end{equation}
Hence, for identical monomers, it follows that
\begin{equation}\label{pperTIdMonomers}
\frac{p}{T} = \frac{2|t^{bp}|}{h}.
\end{equation}

For a dimer consisting of two identical monomers with purine on purine and pyrimidine on pyrimidine,
as mentioned above, the period of $|A_{\mu}(t)|^2, \mu = 1, 2$ is given by Eq.~\ref{periodIdMonomers}.
%In Fig.~\ref{holeselectronsAitsquaredpuonpu}, in the supporting material (Section~\ref{sec:support}), $|A_{\mu}(t)|^2, \mu = 1, 2$ are depicted as the carrier is transferred periodically from base-pair 1 to base-pair 2.
%{\it Hole transfer:}
%for the GG $\equiv$ CC dimer $ T = $ 20.6783 fs, for the AA $\equiv$ TT dimer $ T = $ 103.3917 fs.
%{\it Electron transfer:}
%for the GG $\equiv$ CC dimer $T = $ 103.3917 fs, for the AA $\equiv$ TT dimer $T = $ 71.3046 fs.
In all cases, the maximum transfer percentage, $ p = 1$ (100\%).
For a dimer consisting of two identical monomers with purine on pyrimidine and vice versa,
again, we use Eq.~\ref{periodIdMonomers}.
%{\it Hole transfer:} for the GC dimer $ T = $ 206.7834 fs, for the CG dimer $ T = $ 41.3567 fs,
%for the AT dimer $ T = $  59.0810 fs, for the TA dimer $ T = $ 41.3567 fs.
%{\it Electron transfer:} for the GC dimer $ T =$ 206.7834 fs, for the CG dimer $ T = $ 258.4792 fs,
%for the AT dimer $ T = $ 4135.6675 fs, for the TA dimer $ T = $ 1033.9168 fs.
%In Fig.~\ref{holesAitsquaredpuonpy} and Fig.~\ref{electronsAitsquaredpuonpy}, in the supporting material (Section~\ref{sec:support}), $|A_{\mu}(t)|^2, \mu = 1, 2$ are depicted as the hole and the electron is transferred periodically from base-pair 1 to base-pair 2 in the GC, CG, AT, TA dimers.
In all cases, the maximum transfer percentage, $ p = 1$ (100\%).
Conclusively, for all cases of a dimer consisting of two identical monomers, the maximum transfer percentage $ p = 1$ (100\%).

For a dimer made up of two different monomers we use Eq.~\ref{perioddimers}.
%For holes, the situation is shown in Fig.~\ref{holesAitsquareddimon}, in the supporting material (Section~\ref{sec:support}).
%{\it Hole transfer}:
%for the GA $\equiv$ TC dimer $ T = $ 11.1167 fs with $ p \approx $ 0.3497,
%for the CT $\equiv$ AG dimer $ T = $ 13.5179 fs with $ p \approx $ 0.0385,
%for the GT $\equiv$ AC dimer $ T = $ 13.7550 fs with $ p \approx $ 0.0044,
%for the CA $\equiv$ TG dimer $ T = $ 13.7550 fs with $ p \approx $ 0.0044.
In contrast to the case of identical monomers the maximum transfer percentage $ p < 1$.
For holes, when purines are crosswise to pyrimidines (GT $\equiv$ AC and CA $\equiv$ TG dimers) the maximum transfer percentage is negligible. 
% hence, we expect that insertion of these dimers in a sequence of DNA base-pairs will disrupt {\it hole} transfer.
% an hypothesis that has to be checked later on.  %[$\clubsuit$]
Also AG $\equiv$ CT has very small $p$.
%[{\color{red}$\clubsuit$}]
%[{\color{green}$\clubsuit$}]
%For electrons, the situation is shown in Fig.~\ref{electronsAitsquareddimon}, in the supporting material (Section~\ref{sec:support}).
%{\it Electron transfer}:
%for the GA $\equiv$ TC dimer $T = $ 10.3390 fs with $ p \approx $ 2.5$\times$10$^{-5}$,
%for the CT $\equiv$ AG dimer $T = $ 10.3380 fs with $ p \approx $ 2.2$\times$10$^{-4}$,
%for the GT $\equiv$ AC dimer $T = $ 10.2093 fs with $ p \approx $ 0.0250,
%for the CA $\equiv$ TG dimer $T = $ 10.3020 fs with $ p \approx $ 0.0072.
For electrons, again, we observe that in contrast to the case of identical monomers the maximum transfer percentage $ p < 1$.
Additionally, generally, electrons have smaller maximum transfer percentages than holes.
However, in contrast to the cases of holes, mentioned just above,
when purines are NOT crosswise to pyrimidines (GA $\equiv$ TC and CT $\equiv$ AG) the maximum transfer percentage is negligible. 
% hence, we expect that insertion of these dimers in a sequence of DNA base-pairs will disrupt {\it electron} transfer.
% an hypothesis that has to be checked later on.  [$\maltese$].
Generally, in cases of different monomers $T$ is smaller than
in cases of identical monomers due to difference
between Eq.~\ref{perioddimers} and Eq.~\ref{periodIdMonomers}, i.e.
the extra term containing $\Delta^{bp} = |E^{bp1} - E^{bp2}|$.
Overall, carrier transfer is more difficult for different monomers compared to identical monomers.

%The period $T$, the maximum transfer percentage $p$, and the \textit{pure} maximum transfer rate $\frac{p}{T}$ are depicted in Fig.~\ref{TppperT}, in the supporting material (Section~\ref{sec:support}).

\subsubsection{Dimer \textit{pure} mean transfer rates and other quantities}
\label{dimersquantities}
\begin{figure*}[]
\hspace{0cm}
\centering
\includegraphics[width=8cm]{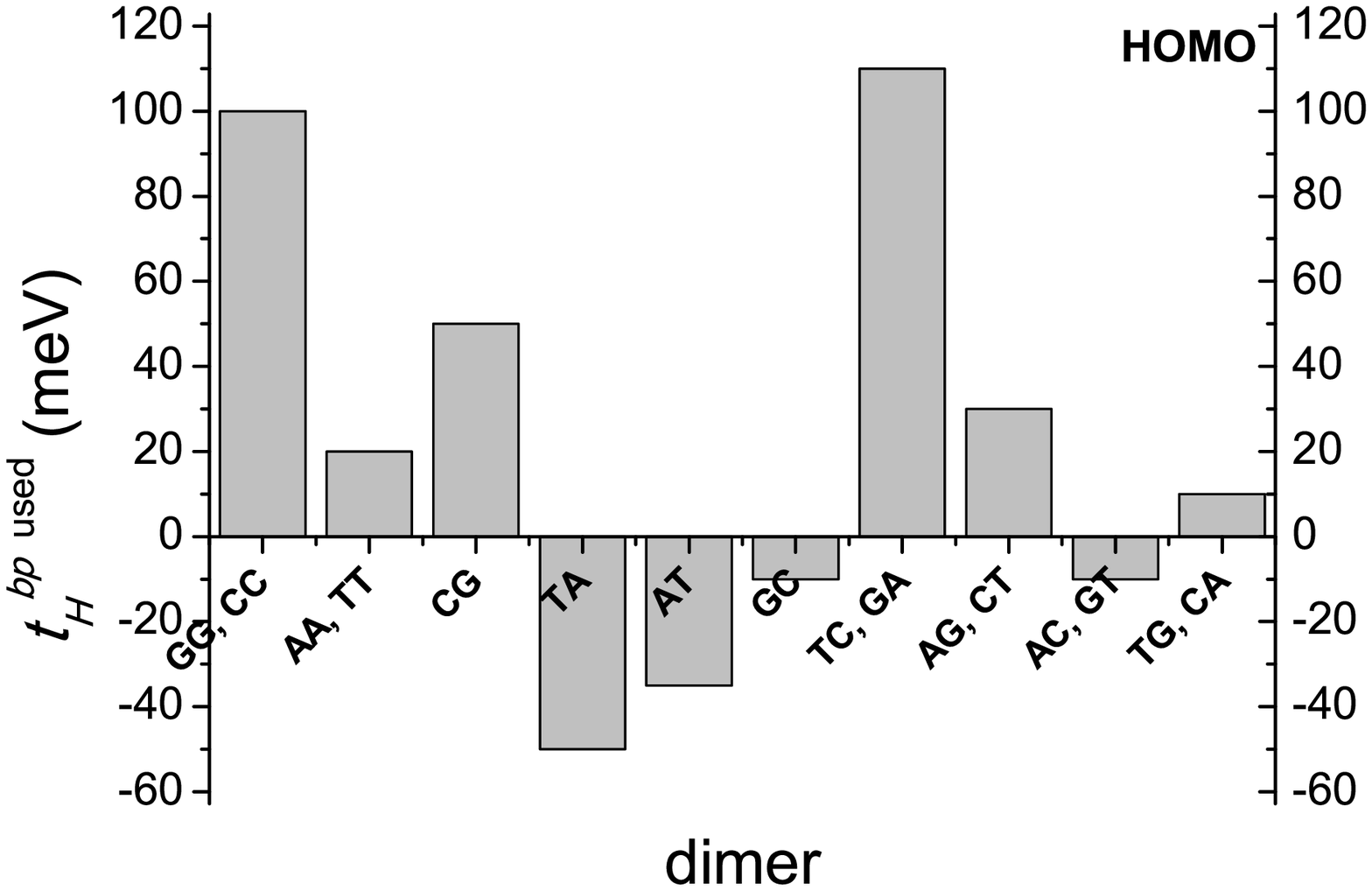}
\includegraphics[width=8cm]{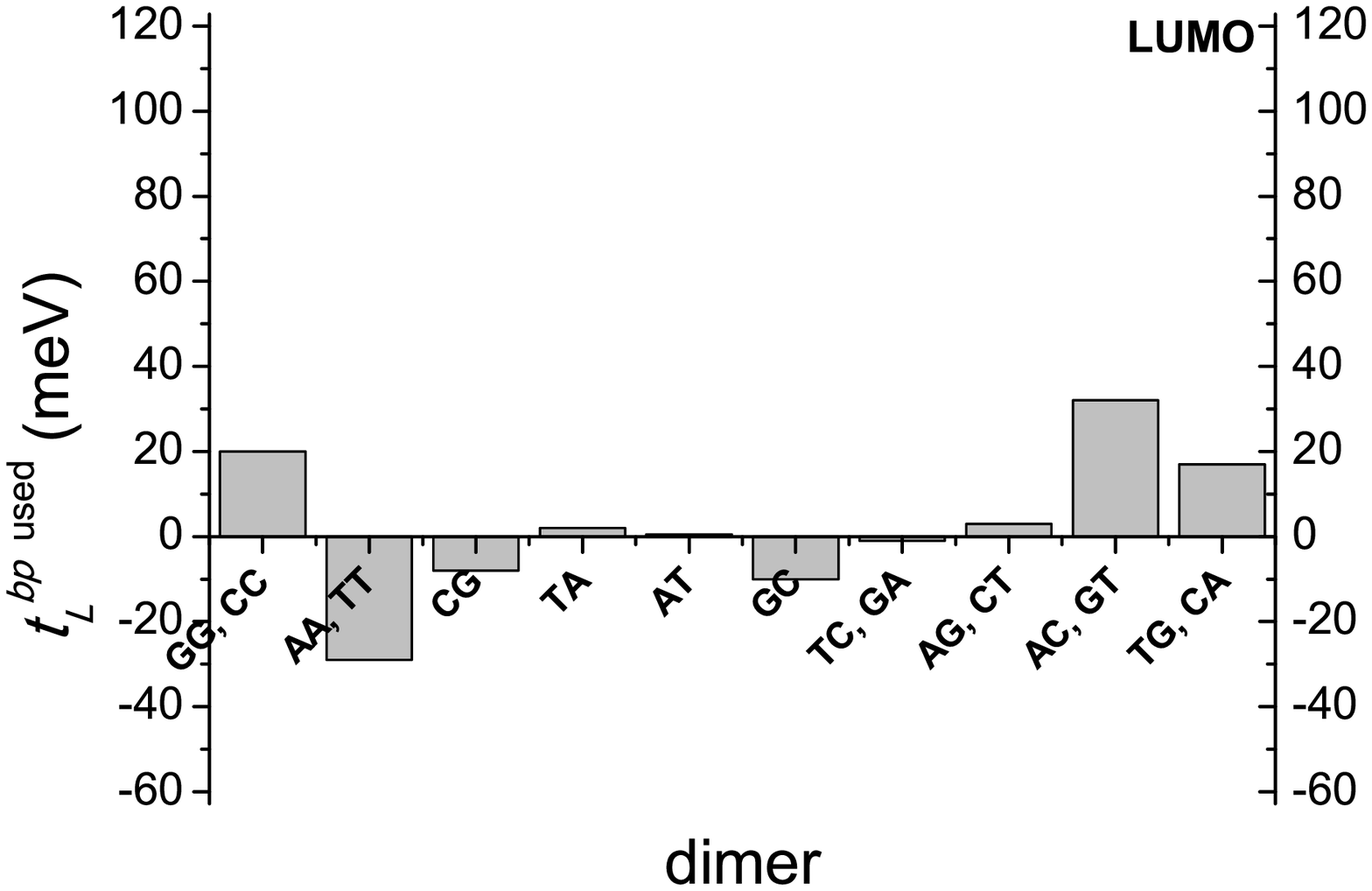}
\includegraphics[width=8cm]{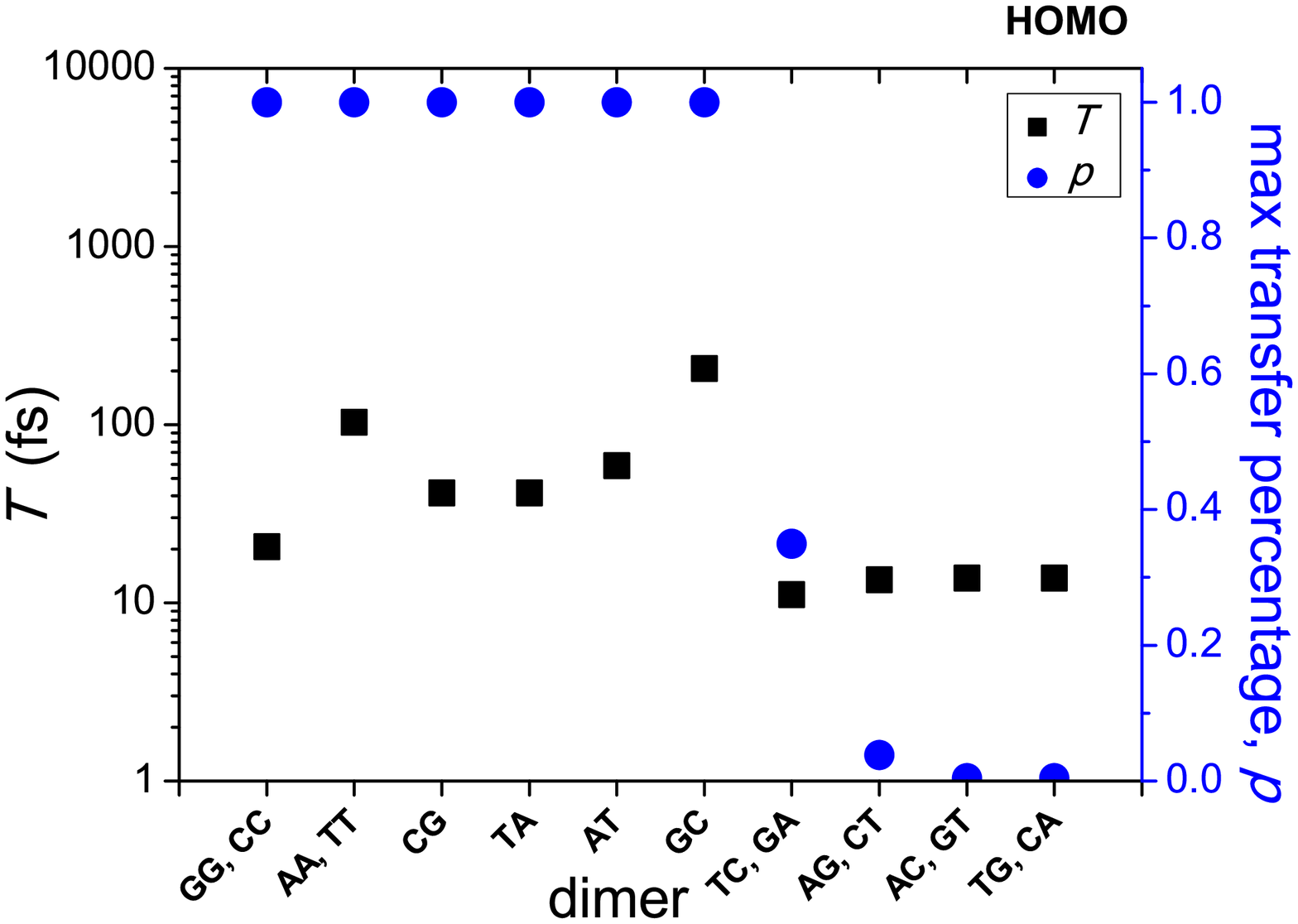}
\includegraphics[width=8cm]{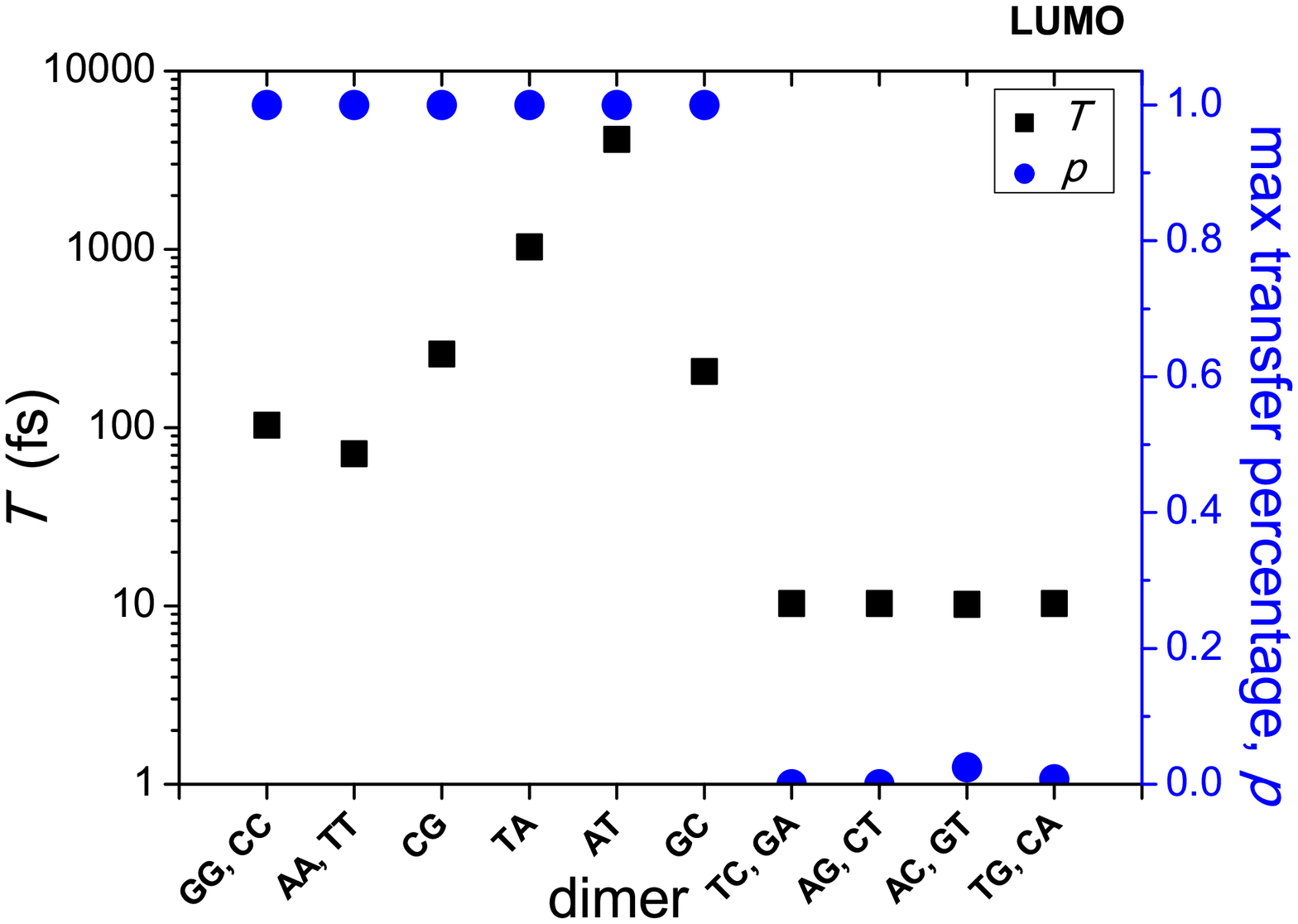}
\includegraphics[width=8cm]{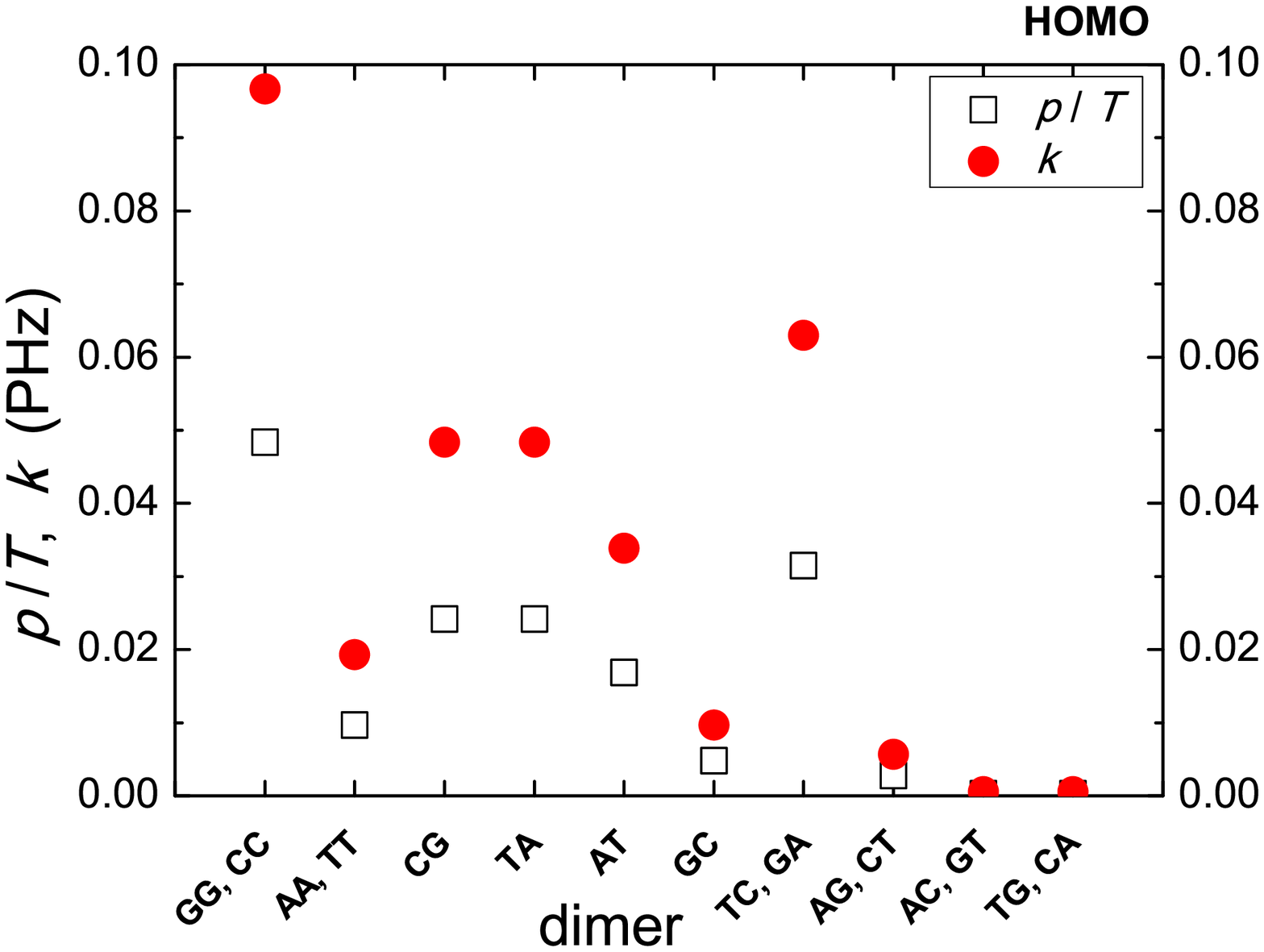}
\includegraphics[width=8cm]{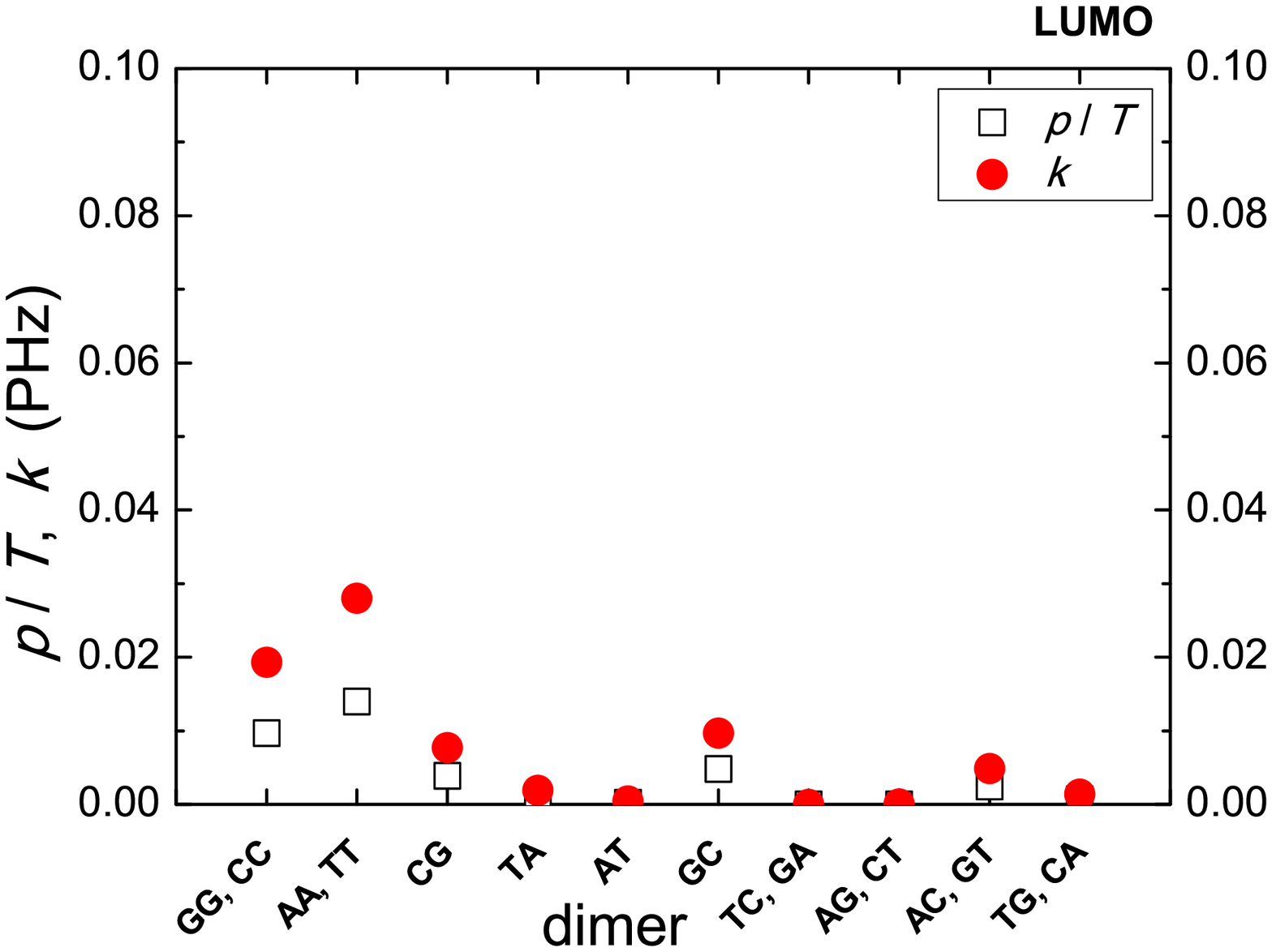}
\includegraphics[width=8cm]{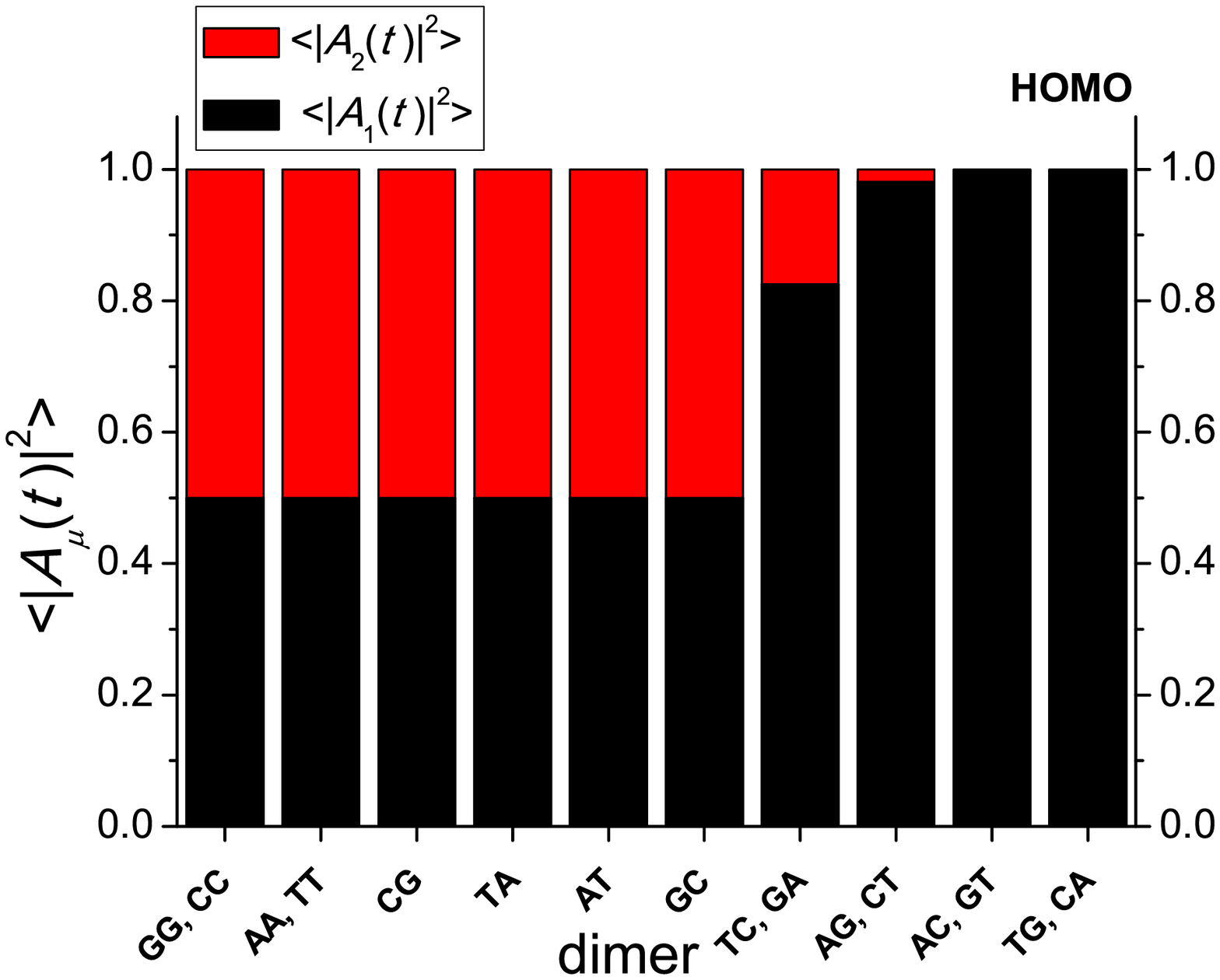}
\includegraphics[width=8cm]{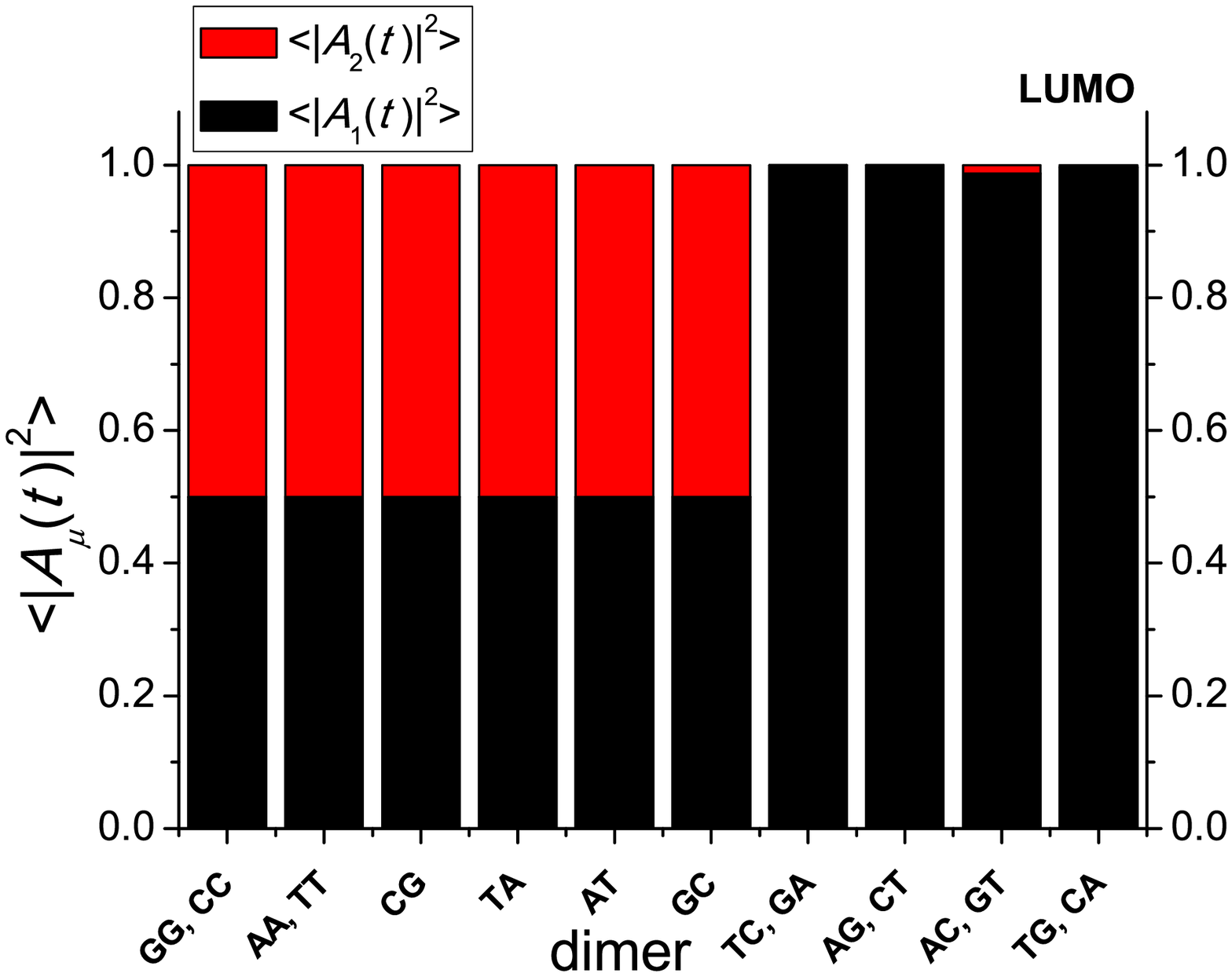}
\caption{Periodic carrier transfer in a base-pair dimer for holes (left column) and electrons (right column).
The hopping integrals $t_{H/L}^{bp \; \textrm{used}}$ (meV) [1st row],
the period of carrier transfer between monomers $T$ (fs) and the maximum transfer percentage $p$ [2nd row],
the \textit{pure} maximum transfer rate defined as $p/T$ (PHz) and
the \textit{pure} mean transfer rate defined as $k = \langle |A_{2}(t)|^2 \rangle / {t_{2}}_{mean} $ (PHz) [3rd row], as well as
$\langle|A_{\mu}(t)|^2\rangle, \; \mu = 1, 2$,
which describe the spread of the carrier over the monomers constituting the dimer [4th row], are shown.
For the dimers made up of identical monomers $p=1$ whereas for the dimers made up of different monomers $p<1$.
In the latter case, the \textit{pure} maximum transfer rate and the \textit{pure} mean transfer rate are negligible
for HOMO hole transfer when purines are crosswise to pyrimidines (GT $\equiv$ AC and CA $\equiv$ TG dimers) and
for LUMO electron transfer when purines are on top of pyrimidines (GA $\equiv$ TC and CT $\equiv$ AG dimers).
For dimers $k = 2 \frac{p}{T}$.}
\label{dimersholeselectrons}
\end{figure*}

If initially i.e. for $t = 0$ we place the carrier at the first monomer (Eq.~\ref{x0}), then $|A_{1}(0)|^2 = 1$, $|A_{2}(0)|^2 = 0$.
Hence, a \textit{pure} mean transfer rate can be defined as
\begin{equation}\label{meantransferrate2}
k = \frac{\langle |A_{2}(t)|^2 \rangle}{{t_{2}}_{mean}},
\end{equation}
where ${t_{2}}_{mean}$ is the first time $|A_{2}(t)|^2$ becomes equal to $\langle |A_{2}(t)|^2 \rangle$ i.e.
``the mean transfer time''. This definition as well as the analogous definitions for trimers (Eq.~\ref{meantransferrate3}) and polymers (Eq.~\ref{meantransferrateN}) take into account not only the transfer time but also the mean magnitude of charge transfer expressed by
$\langle|A_{\mu}(t)|^2\rangle$.
Figure~\ref{dimersholeselectrons} shows quantities relevant to the periodic carrier transfer in a base-pair dimer, specifically,
the hopping integrals $t_{H/L}^{bp \; \textrm{used}}$,
the period of carrier transfer between monomers $T$ and the maximum transfer percentage $p$,
the \textit{pure} maximum transfer rate defined as $p/T$ and
the \textit{pure} mean transfer rate defined as $k = \langle |A_{2}(t)|^2 \rangle / {t_{2}}_{mean} $, as well as
$\langle|A_{\mu}(t)|^2\rangle, \; \mu = 1, 2$,
which describe the spread of the carrier over the monomers constituting the dimer.
For the dimers made up of identical monomers $p=1$ whereas for the dimers made up of different monomers $p<1$.
In the latter case, the \textit{pure} maximum transfer rate and the \textit{pure} mean transfer rate are negligible
for HOMO hole transfer when purines are crosswise to pyrimidines (GT $\equiv$ AC and CA $\equiv$ TG dimers) and
for LUMO electron transfer when purines are on top of pyrimidines (GA $\equiv$ TC and CT $\equiv$ AG dimers).
For dimers $k = 2 \frac{p}{T}$.

%%%%%%%%%%%%%%%%%%%%%%%%%%
\subsection{Trimers}%%%%%%%
\label{subsec:trimers}%%%%%%
%%%%%%%%%%%%%%%%%%%%%%%%%%
For any trimer, the solution of our problem is
\begin{equation}\label{trimersolution}
\vec{x}(t) = \sum_{k=1}^{3} c_k \vec{v_k} e^{-\frac{i}{\hbar} \lambda_k t}.
\end{equation}
Let us suppose further that $\lambda_1 \le \lambda_2 \le \lambda_3$.
We are interested in the quantities $|A_{\mu}(t)|^2, \mu = 1, 2, 3$ (cf. Eq.~\ref{psi-total} and Eq.~\ref{x}) since they provide the probabilities of finding the carrier at the base-pair $\mu$. From Eq.~\ref{trimersolution}, we conclude that $|A_{\mu}(t)|^2, \mu = 1, 2, 3$ are sums of terms containing constants and periodic functions with periods
\begin{equation}\label{periods}
T_{21} = \frac{h}{\lambda_2 - \lambda_1}, \\
T_{32} = \frac{h}{\lambda_3 - \lambda_2}, \\
T_{31} = \frac{h}{\lambda_3 - \lambda_1}.
\end{equation}
In case of double degeneracy, e.g. $\lambda_2 = \lambda_1 \Rightarrow T_{21} \to \infty$, i.e. the terms containing $T_{21}$ are constant and the only period that remains is $T_{32} = T_{31}$, hence $|A_{\mu}(t)|^2, \mu = 1, 2, 3$ are periodic. In case of triple degeneracy, $\lambda_3 = \lambda_2 = \lambda_1$ all periods are  infinite, hence $|A_{\mu}(t)|^2, \mu = 1, 2, 3$ are constants.

\subsubsection{A trimer consisting of identical monomers}
\label{trimersidmon}
There are 8 such trimers,
4 based on G-C monomers [GGG $\equiv$ CCC (0 times crosswise purines), GGC $\equiv$ GCC (1 time crosswise purines), CGG $\equiv$ CCG (1 time crosswise purines), GCG $\equiv$ CGC (2 times crosswise purines)], and
4 based on A-T monomers [AAA $\equiv$ TTT (0 times crosswise purines), AAT $\equiv$ ATT (1 time crosswise purines), TAA $\equiv$ TTA (1 time crosswise purines), ATA $\equiv$ TAT (2 times crosswise purines)].

In the cases of 0 times crosswise purines
\begin{equation}\label{ATrimersIdMonomers0}
\textrm{A} = \left[
\begin{array}{ccc}
E^{bp} & t^{bp} & 0      \\
t^{bp} & E^{bp} & t^{bp} \\
0      & t^{bp} & E^{bp} \end{array} \right]
\end{equation}
with eigenvalues
\begin{equation}\label{ETrimersIdMonomers0}
\lambda_2 = E^{bp}, \;\;\;\;\;
\lambda_{1,3} = E^{bp} \mp t^{bp} \sqrt{2}.
\end{equation}
Hence, two periods are involved in $|A_{\mu}(t)|^2, \mu = 1, 2, 3$:
\begin{equation}\label{periodsTrimersIdMonomers0}
T_{M} = \frac{h}{t^{bp} \sqrt{2}}, \;\;\;\;\;
T_{E} = \frac{h}{2 t^{bp} \sqrt{2}} \;\;
\Rightarrow \;\; \frac{T_{M}}{T_{E}} = \frac{2}{1}.
\end{equation}
$T_{M}=T_{M(21)}=T_{M(32)}$ involves the Medium eigenvalue, $T_{E}=T_{E(31)}$ involves only the Edge eigenvalues.
Since $\frac{T_{M}}{T_{E}} = \frac{2}{1}$ it follows that $|A_{\mu}(t)|^2, \mu = 1, 2, 3$ are periodic.

In the cases of 1 or 2 times crosswise purines
\begin{equation}\label{ATrimersIdMonomers12}
\textrm{A} = \left[
\begin{array}{ccc}
E^{bp} & t^{bp}  & 0      \\
t^{bp} & E^{bp}  & t^{bp'} \\
0      & t^{bp'} & E^{bp} \end{array} \right]
\end{equation}
with eigenvalues
\begin{equation}\label{ETrimersIdMonomers12}
\lambda_2 = E^{bp}, \;\;\;\;\;
\lambda_{1,3} = E^{bp} \mp \sqrt{{t^{bp}}^2+{t^{bp'}}^2}.
\end{equation}
Hence, two periods are involved in $|A_{\mu}(t)|^2, \mu = 1, 2, 3$:
\begin{equation}\label{periodsTrimersIdMonomers12}
T_{M} = \frac{h}{\sqrt{{t^{bp}}^2+{t^{bp'}}^2}},
T_{E} = \frac{h}{2 \sqrt{{t^{bp}}^2+{t^{bp'}}^2}}
\Rightarrow \frac{T_{M}}{T_{E}} = \frac{2}{1}.
\end{equation}
Again, since $\frac{T_{M}}{T_{E}} = \frac{2}{1}$ it follows that $|A_{\mu}(t)|^2, \mu = 1, 2, 3$ are periodic.

Conclusively, in all cases of a trimer consisting of identical monomers, $|A_{\mu}(t)|^2, \mu = 1, 2, 3$ are periodic with period $T_{M}$.

\subsubsection{A trimer consisting of different monomers}
\label{trimersdimon}
Suppose that we have a trimer consisting of different monomers. There are 24 different such trimers:
GGA $\equiv$ TCC, GGT $\equiv$ ACC, GCA $\equiv$ TGC, GCT $\equiv$ AGC, GAG $\equiv$ CTC, GAC $\equiv$ GTC,
GAA $\equiv$ TTC, GAT $\equiv$ ATC, GTG $\equiv$ CAC, GTA $\equiv$ TAC, GTT $\equiv$ AAC, CGA $\equiv$ TCG,
CGT $\equiv$ ACG, CCA $\equiv$ TGG, CCT $\equiv$ AGG, CAG $\equiv$ CTG, CAA $\equiv$ TTG, CAT $\equiv$ ATG,
CTA $\equiv$ TAG, AGA $\equiv$ TCT, AGT $\equiv$ ACT, ACA $\equiv$ TGT, TGA $\equiv$ TCA, CTT $\equiv$ AAG.
For example, suppose that we refer to HOMO charge transfer in GAC $\equiv$ GTC, then
\begin{equation}\label{ATrimersDiMonomersGAC}
\textrm{A} = \left[
\begin{array}{ccc}
E^{bp} & t^{bp}  & 0      \\
t^{bp} & E^{bp''}  & t^{bp'} \\
0      & t^{bp'} & E^{bp} \end{array} \right]
\end{equation}
where $E^{bp''} > E^{bp}$, and the eigenvalues are
\begin{eqnarray}
\lambda_2 = E^{bp}, \\ \nonumber
\lambda_{1,3} \! = \! \frac{E^{bp}+E^{bp''}}{2} \! \mp \!
\sqrt{\left(\frac{E^{bp}-E^{bp''}}{2}\right)^2 \! + \! {t^{bp}}^2+{t^{bp'}}^2 }.
\label{ETrimersIdMonomers12}
\end{eqnarray}
Hence, three periods are involved in $|A_{\mu}(t)|^2, \mu = 1, 2, 3$.
Defining $\Delta^{bp} = |E^{bp} - E^{bp''}|$,
\begin{eqnarray}\label{periodsTrimersIdMonomers12}
T_{M(32)} = \frac{h}{\frac{\Delta^{bp}}{2} \! + \!
\sqrt{\frac{{\Delta^{bp}}^2}{4} \! + \! {t^{bp}}^2+{t^{bp'}}^2 }}, \\ \nonumber
T_{E(31)} = \frac{h}{2 \sqrt{\frac{{\Delta^{bp}}^2}{4} \! + \! {t^{bp}}^2+{t^{bp'}}^2 }}, \\ \nonumber
T_{M(21)} = \frac{h}{-\frac{\Delta^{bp}}{2} \! + \!
\sqrt{\frac{{\Delta^{bp}}^2}{4} \! + \! {t^{bp}}^2+{t^{bp'}}^2 }}
\end{eqnarray}
Therefore, $\frac{T_{M(32)}}{T_{E(31)}}$ and $\frac{T_{M(21)}}{T_{E(31)}}$ may be irrational numbers, hence  $|A_{\mu}(t)|^2, \mu = 1, 2, 3$ may be non periodic.

\subsubsection{Trimer \textit{pure} mean transfer rates and other quantities}
\label{trimersquantities}
\begin{figure*}[]
\hspace{0cm}
\centering
\includegraphics[width=16cm]{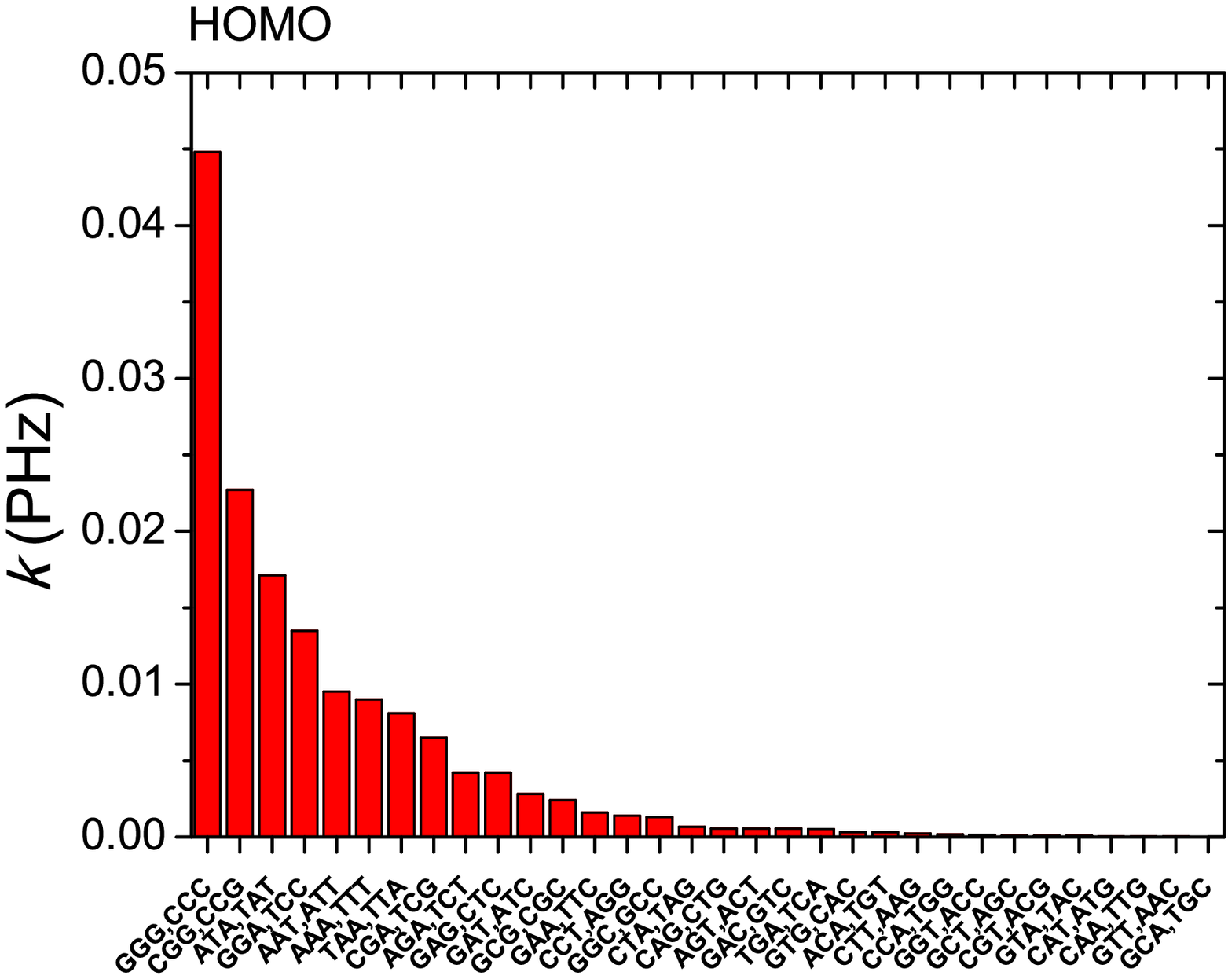}
\includegraphics[width=16cm]{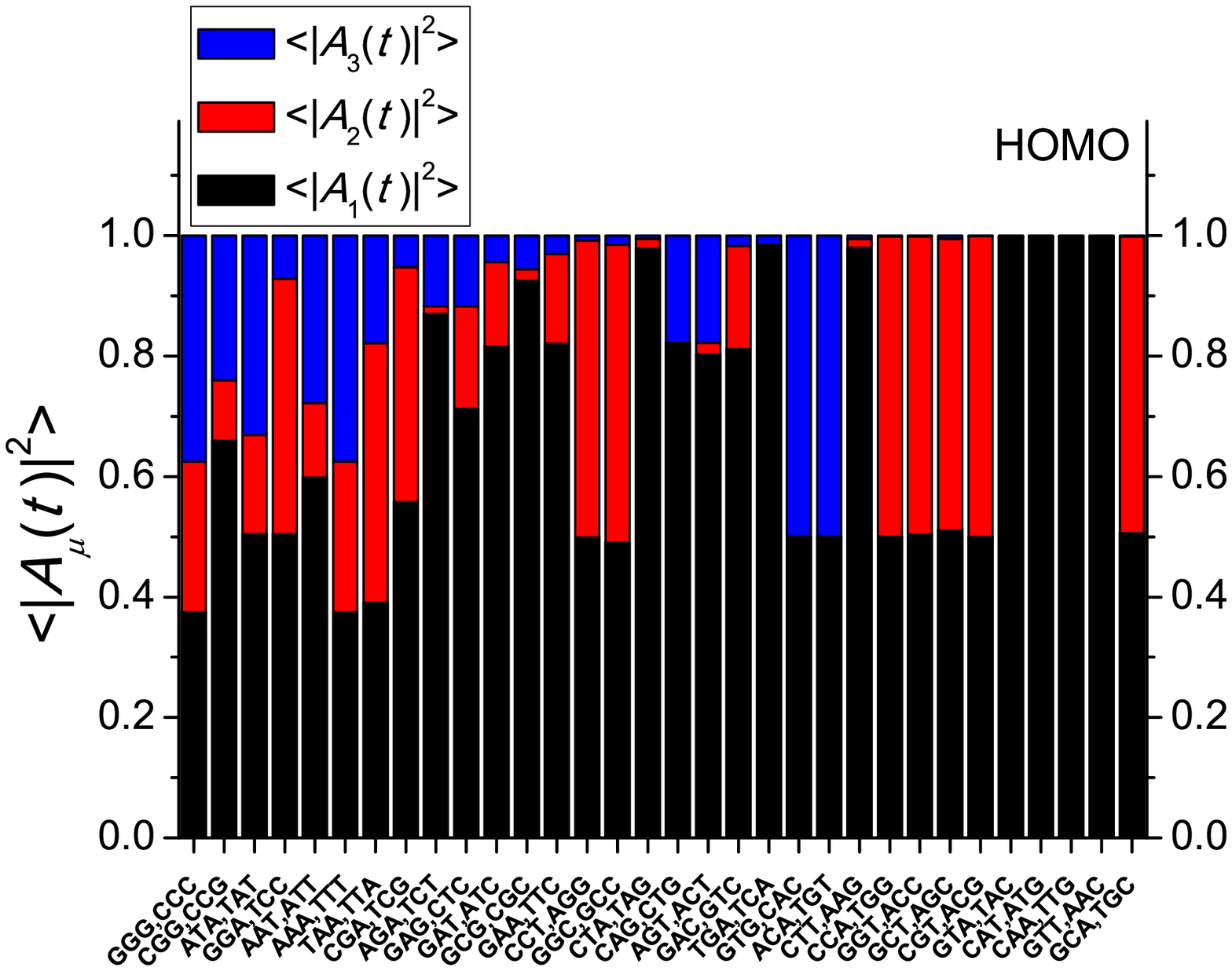}
\caption{The HOMO \textit{pure} mean transfer rate $k$ for all possible trimers as well as
$\langle|A_{\mu}(t)|^2\rangle, \; \mu = 1, 2, 3$, which describe the spread of the hole over the monomers constituting the trimer.
The trimers are shown in descending HOMO $k$ order. Initially, we place the hole in monomer 1.}
\label{trimersholes}
\end{figure*}

The calculations show that for trimers consisting of identical monomers $k \approx 1.3108 \frac{p}{T}$.
However, since for trimers consisting of different monomers $|A_{\mu}(t)|^2, \mu = 1, 2, 3$ may be non periodic, from now on
we will use only the \textit{pure} mean transfer rate $k$.
Assuming that initially i.e. for $t = 0$ we place the carrier at the first monomer (Eq.~\ref{x0}), then
$|A_{1}(0)|^2 = 1$, $|A_{2}(0)|^2 = 0$, $|A_{3}(0)|^2 = 0$. Hence, a \textit{pure} mean transfer rate can be defined as
\begin{equation}\label{meantransferrate3}
k = \frac{\langle |A_{3}(t)|^2 \rangle}{{t_{3}}_{mean}},
\end{equation}
where ${t_{3}}_{mean}$ is the first time $|A_{3}(t)|^2$ becomes equal to $\langle |A_{3}(t)|^2 \rangle$
i.e. ``the mean transfer time''.

The HOMO \textit{pure} mean transfer rate $k$ for all possible trimers as well as $\langle|A_{\mu}(t)|^2\rangle, \; \mu = 1, 2, 3$, which describe the spread of the hole over the monomers constituting the trimer are shown in Figure~\ref{trimersholes}.
As expected, $k$ is very small when trimers include dimers with very small $k$, primarily purines crossswise to pyrimidines (GT $\equiv$ AC, CA $\equiv$ TG), secondarily AG $\equiv$ CT, and thirdly GC. We underline that in some cases $k$ is very small due to the very large value of the mean transfer time although \textit{finally} the carrier may have an appreciable or even remarkable occupancy of the last monomer. For example, in GTG$\equiv$CAC and in ACA$\equiv$TGT, $|A_{3}(t)|^2 \approx 0.4989$, although $k \approx 3.2\times10^{-4}$. This remark, displayed here for trimers, holds also for polymers and one should not forget it when comparing with results from different experimental techniques.

\begin{figure*}[]
\hspace{0cm}
\centering
\includegraphics[width=16cm]{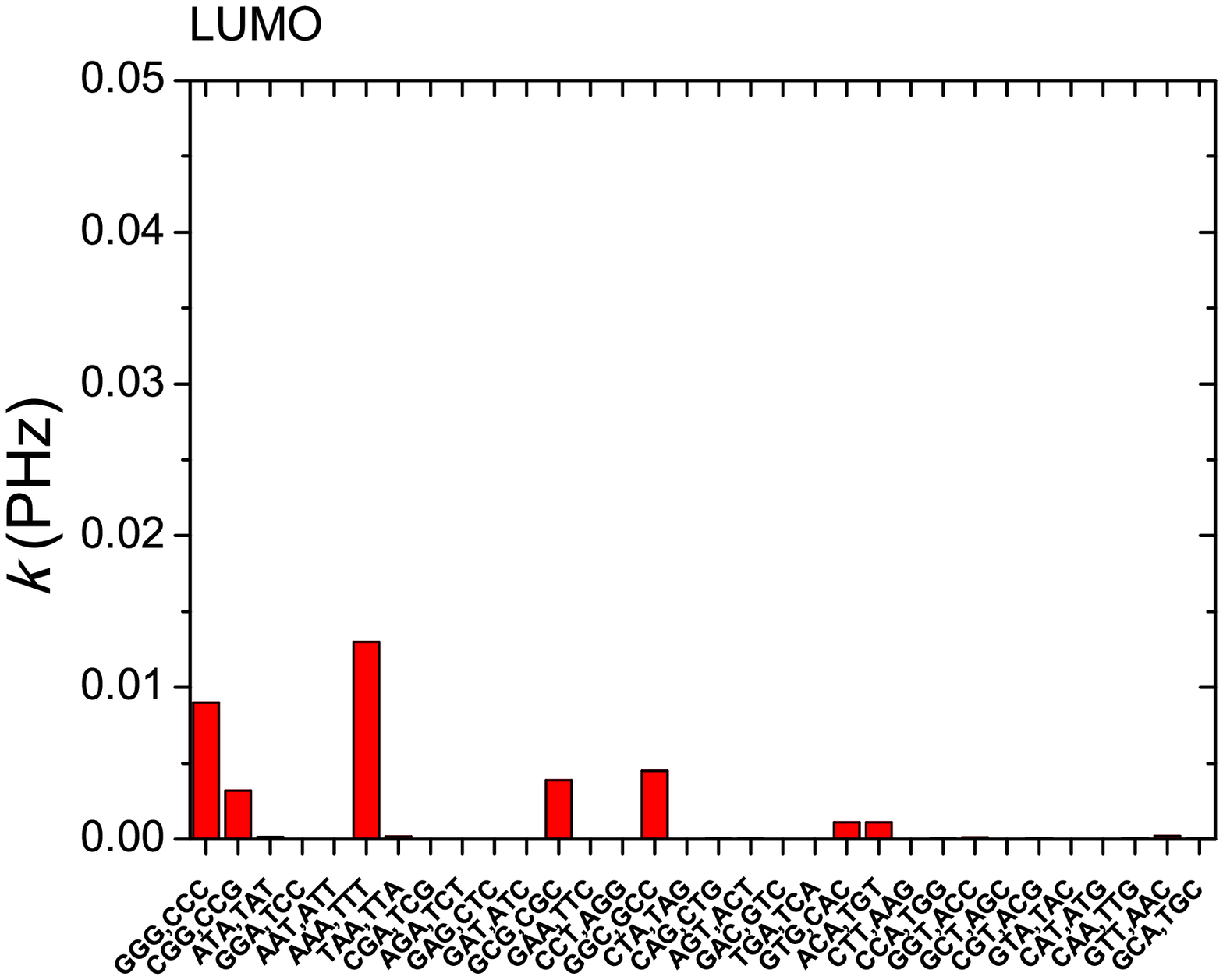}
\includegraphics[width=16cm]{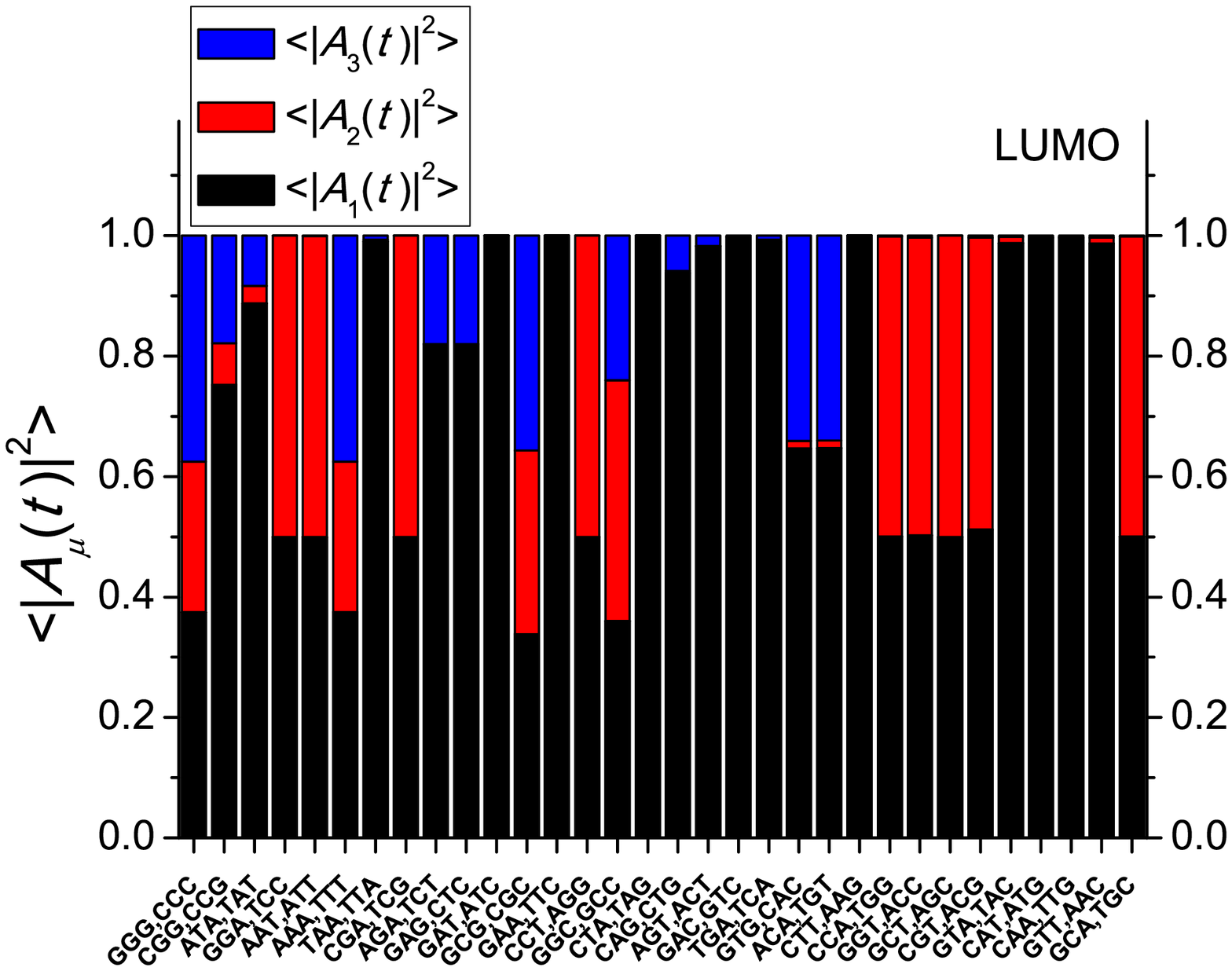}
\caption{The LUMO \textit{pure} mean transfer rate $k$ for all possible trimers as well as
$\langle|A_{\mu}(t)|^2\rangle, \; \mu = 1, 2, 3$, which describe the spread of the electron over the monomers constituting the trimer.
The trimers are still shown in descending HOMO $k$ order, as in Fig.~\ref{trimersholes}. Initially, we place the electron in monomer 1.}
\label{trimerselectrons}
\end{figure*}

The LUMO \textit{pure} mean transfer rate $k$ for all possible trimers as well as $\langle|A_{\mu}(t)|^2\rangle, \; \mu = 1, 2, 3$, which describe the spread of the electron over the monomers constituting the trimer are shown in Figure~\ref{trimerselectrons}.

\label{trimersquantities}

%%%%%%%%%%%%%%%%%%%%%%%%%%%%%%
\subsection{Polymers}%%%%%%%%
\label{subsec:polymers}%%%%%%
%%%%%%%%%%%%%%%%%%%%%%%%%%%%%%
Supposing that initially i.e. for $t = 0$ we place the carrier at the first monomer (Eq.~\ref{x0}), then $|A_{1}(0)|^2 = 1$, while all other $|A_{j}(0)|^2 = 0$, $j=2, \dots, N$. Hence, for a polymer consisting of $N$ monomers, a \textit{pure} mean transfer rate can be defined as
\begin{equation}\label{meantransferrateN}
k = \frac{\langle |A_{N}(t)|^2 \rangle}{{t_{N}}_{mean}},
\end{equation}
where ${t_{N}}_{mean}$ is the first time $|A_{N}(t)|^2$ becomes equal to $\langle |A_{N}(t)|^2 \rangle$ i.e.
``the mean transfer time''.

Below, increasing the number of base-pairs or monomers $N$, we study various characteristic polymers:
poly(dG)-poly(dC), poly(dA)-poly(dT), GCGCGC..., CGCGCG..., ATATAT..., TATATA...
as well as DNA segments that have been experimentally studied in the past.
As usual, we only write down the 5'- 3' DNA strand.
If we fit $k(d)$ --i.e. the \textit{pure} mean transfer rate $k$ as a function of the charge transfer distance $d = N \times$ 3.4 {\AA}--
exponentially, as $k = k_0 \textrm{exp}(-\beta d)$, we obtain an estimation of $k_0$ and of the distance dependence parameter or inverse decay length $\beta$~\cite{Marcus}. These quantities are displayed in Table~\ref{table:ExponentialkofdFit}.
If, instead, we fit $k(N)$ --i.e. the \textit{pure} mean transfer rate $k$ as a function of the number of monomers $N$-- in a power law, as $k = k_0' N^{-\eta}$, we obtain an estimation of $k_0'$ and $\eta$.
These quantities are displayed in Table~\ref{table:PowerkofNFit}.
Values of $\beta$, in the range $\approx$ 0.3-1.5 {\AA}$^{-1}$, for various compounds, have been displayed in the literature at least 30 years now, see e.g Table IV of Ref.~\cite{Marcus}. In Table~\ref{table:ExponentialkofdFit} the values of $\beta$ are in the range $\approx$ 0.2-2 {\AA}$^{-1}$,
with smaller values for periodic polymers like ATATAT..., poly(dG)-poly(dC), poly(dA)-poly(dT). However, for efficient charge transfer, a small value of $\beta$ is not enough; one should also take into account the magnitude of $k_0$. The values of $k_0$ assumed in Ref.~\cite{Marcus} are 10$^{-2}$-10$^{-1}$ PHz which coincides with most of the $k_0$ values shown in Table~\ref{table:ExponentialkofdFit}, although generally, the values of $k_0$ fall in the wider range $\approx 10^{-4}$-10 PHz.
For the power law fit, $\eta \approx$ 1.7 - 17; most of the $k_0'$ values shown in Table~\ref{table:PowerkofNFit} are in the range $\approx 10^{-2}$-10$^{-1}$ PHz, although generally, the values of $k_0'$ fall in the wider range $\approx 10^{-4}$-10$^3$ PHz.
The $\beta$-value for charge transfer from an initial site (donor) to a final site (acceptor) depends on the mediating molecules, the so-called bridge. From Table~\ref{table:ExponentialkofdFit} we conclude that there are no universal (even approximately) values of $\beta$ and $k_0$ for DNA, instead, each specific DNA segment is unique and one should use an efficient and easy way to predict $\beta$ and $k_0$ of each DNA segment under investigation. It is hoped that the present  work will contribute in this direction.
$\beta$ values for different systems include
$\approx$ 1.0 - 1.4 {\AA}$^{-1}$ for protein-bridged systems~\cite{WinklerGray:1992,Moser:1992,GrayWinkler:2005},
$\approx$ 1.55 - 1.65 {\AA}$^{-1}$ for aqueous glass bridges~\cite{GrayWinkler:2005},
$\approx$ 0.2 - 1.4 {\AA}$^{-1}$ for DNA segments~\cite{Lewis:1997,Holmlin:1998,Henderson:1999,Wan:2000,KawaiMajima:2013,KalosakasSpanou:2013},
$\approx$ 0.8 - 1.0 {\AA}$^{-1}$ for saturated hydrocarbon bridges~\cite{Johnson:1989,Oevering:1987},
$\approx$ 0.2 - 0.6 {\AA}$^{-1}$ for unsaturated phenylene~\cite{Helms:1992,Ribou:1994}, polyene~\cite{Arrhenius:1986,Effenberger:1991,Tolbert:1992} and polyyne~\cite{Tour:1996,Grosshenny:1996,Sachs:1997} bridges, and much smaller values ($<$ 0.05 {\AA}$^{-1}$), suggesting a molecular-wire-like behavior, for a p-phenylenevinylene bridge~\cite{Davis:1998}.
Hence, it seems that charge transfer in ATATAT..., poly(dG)-poly(dC) and poly(dA)-poly(dT) is almost molecular-wire-like.
Since a carrier can migrate along DNA over 200 {\AA}~\cite{Meggers:1998,Henderson:1999,KawaiMajima:2013},
in the present calculations for polymers $d$ is extending up to 204 {\AA} ($N$ up to 60 base-pairs).

\begin{widetext}

\begin{table}[h!]
\caption{Estimated $k_0$ and $\beta$ of the exponential fit $k = k_0 \textrm{exp}(-\beta d)$ for various DNA polymers.}
\scriptsize{
\begin{tabular}{|l|c|c|c|c|} \hline
DNA segment       & $k_0$ (PHz)       & $\beta$ ({\AA}$^{-1}$) & Correlation &H/L\\
                  &                   &                        & Coefficient &   \\ \hline
poly(dG)-poly(dC) & 0.176 $\pm$ 0.007 & 0.189 $\pm$ 0.008      & 0.988       & H \\ \hline
poly(dG)-poly(dC) & 0.035 $\pm$ 0.001 & 0.189 $\pm$ 0.007      & 0.989       & L \\ \hline
poly(dA)-poly(dT) & 0.035 $\pm$ 0.001 & 0.189 $\pm$ 0.008      & 0.988       & H \\ \hline
poly(dA)-poly(dT) & 0.051 $\pm$ 0.002 & 0.189 $\pm$ 0.008      & 0.989       & L \\ \hline
GCGCGC...         & 0.032 $\pm$ 0.003 & 0.358 $\pm$ 0.023      & 0.988       & H \\ \hline
ATATAT...         & 0.057 $\pm$ 0.002 & 0.168 $\pm$ 0.008      & 0.985       & H \\ \hline
CGCGCG...         & 0.932 $\pm$ 0.233 & 0.871 $\pm$ 0.074      & 0.994       & H \\ \hline
TATATA...         & 0.110 $\pm$ 0.005 & 0.251 $\pm$ 0.012      & 0.985       & H \\ \hline
AGTGCCAAGCTTGCA~\cite{Murphy:1993}& 0.059 $\pm$ 0.002 & 0.685 $\pm$ 0.008      & 1.000       & H \\ \hline
AGTGCCAAGCTTGCA~\cite{Murphy:1993}& $(9.8 \! \pm \! 2.6) \!\! \times \!\! 10^{-5}$& 0.197 $\pm$ 0.059      & 0.808       & L \\ \hline
TAGAGGTGTTATGA~\cite{VariationWan:2000}   & 4.306 $\pm$ 5.001 & 1.321 $\pm$ 0.342      & 0.998       & H \\ \hline
TAGAGGTGTTATGA~\cite{VariationWan:2000}   & 2.877 $\pm$ 0.833 & 2.154 $\pm$ 0.085      & 1.000       & L \\ \hline
\end{tabular} }   \label{table:ExponentialkofdFit}
\end{table}

\begin{table}[h!]
\caption{Estimated $k_0'$ and $\eta$ of the power fit $k = k_0' N^{-\eta}$ for various DNA polymers.}
\scriptsize{
\begin{tabular}{|l|c|c|c|c|} \hline			
DNA segment       & $k_0'$  (PHz)     & $\eta$            & Correlation &H/L\\
                  &                   &                   & Coefficient &   \\ \hline
poly(dG)-poly(dC) & 0.359 $\pm$ 0.001 & 1.893 $\pm$ 0.002 & 1.000       & H \\ \hline
poly(dG)-poly(dC) & 0.072 $\pm$ 0.000 & 1.895 $\pm$ 0.002 & 1.000       & L \\ \hline
poly(dA)-poly(dT) & 0.072 $\pm$ 0.000 & 1.892 $\pm$ 0.002 & 1.000       & H \\ \hline
poly(dA)-poly(dT) & 0.105 $\pm$ 0.000 & 1.893 $\pm$ 0.002 & 1.000       & L \\ \hline
GCGCGC...         & 0.087 $\pm$ 0.008 & 3.176 $\pm$ 0.127 & 0.993       & H \\ \hline
ATATAT...         & 0.117 $\pm$ 0.004 & 1.776 $\pm$ 0.035 & 0.994       & H \\ \hline
CGCGCG...         & 5.082 $\pm$ 1.619 & 6.715 $\pm$ 0.458 & 0.994       & H \\ \hline
TATATA...         & 0.236 $\pm$ 0.007 & 2.295 $\pm$ 0.035 & 0.997       & H \\ \hline
AGTGCCAAGCTTGCA~\cite{Murphy:1993}& 1.383 $\pm$ 0.826 & 4.487 $\pm$ 0.487 & 0.997       & H \\ \hline
AGTGCCAAGCTTGCA~\cite{Murphy:1993}& $(2.2 \! \pm \! 1.0)\!\! \times \!\! 10^{-4}$ & 2.176 $\pm$ 0.543 & 0.761       & L \\ \hline
TAGAGGTGTTATGA~\cite{VariationWan:2000}   &46.300 $\pm$53.288 & 9.902 $\pm$ 1.660 & 0.998       & H \\ \hline
TAGAGGTGTTATGA~\cite{VariationWan:2000}   & 203.457$\pm$99.552& 16.708 $\pm$ 0.706& 1.000       & L \\ \hline
\end{tabular} }   \label{table:PowerkofNFit}
\end{table}

\end{widetext}

\begin{figure*}[]
\centering
\includegraphics[width=8cm]{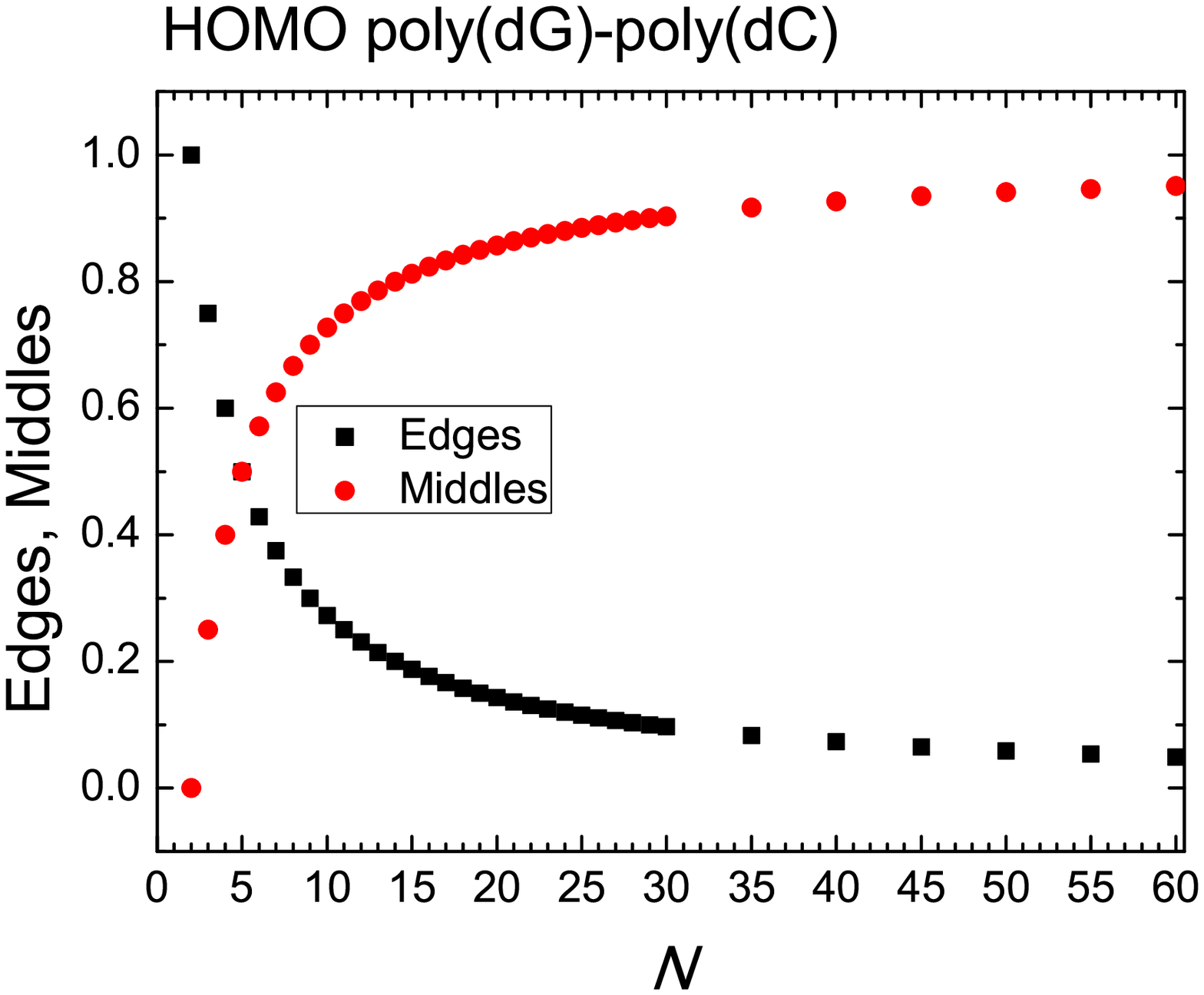}
\includegraphics[width=8cm]{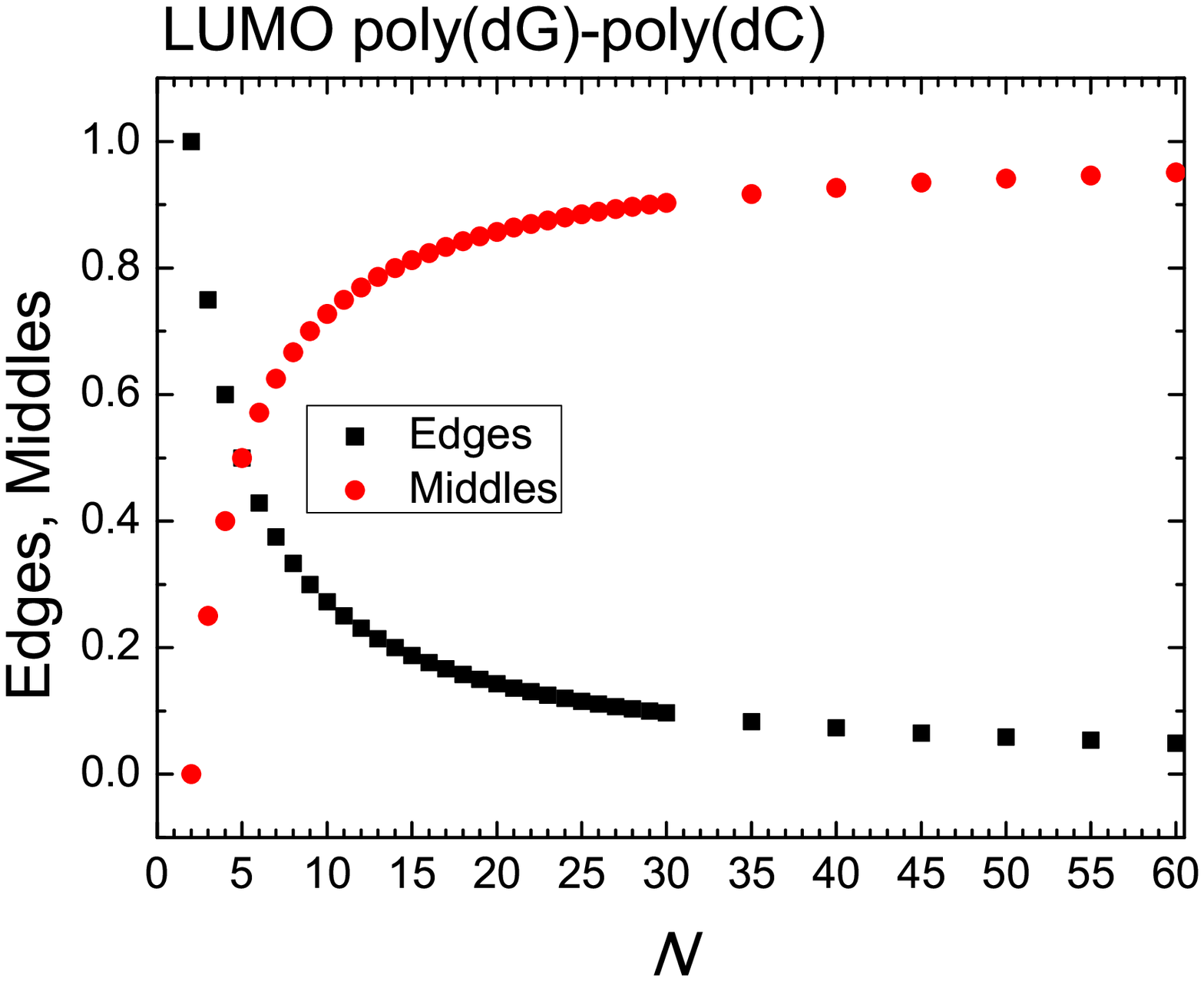}
\includegraphics[width=8cm]{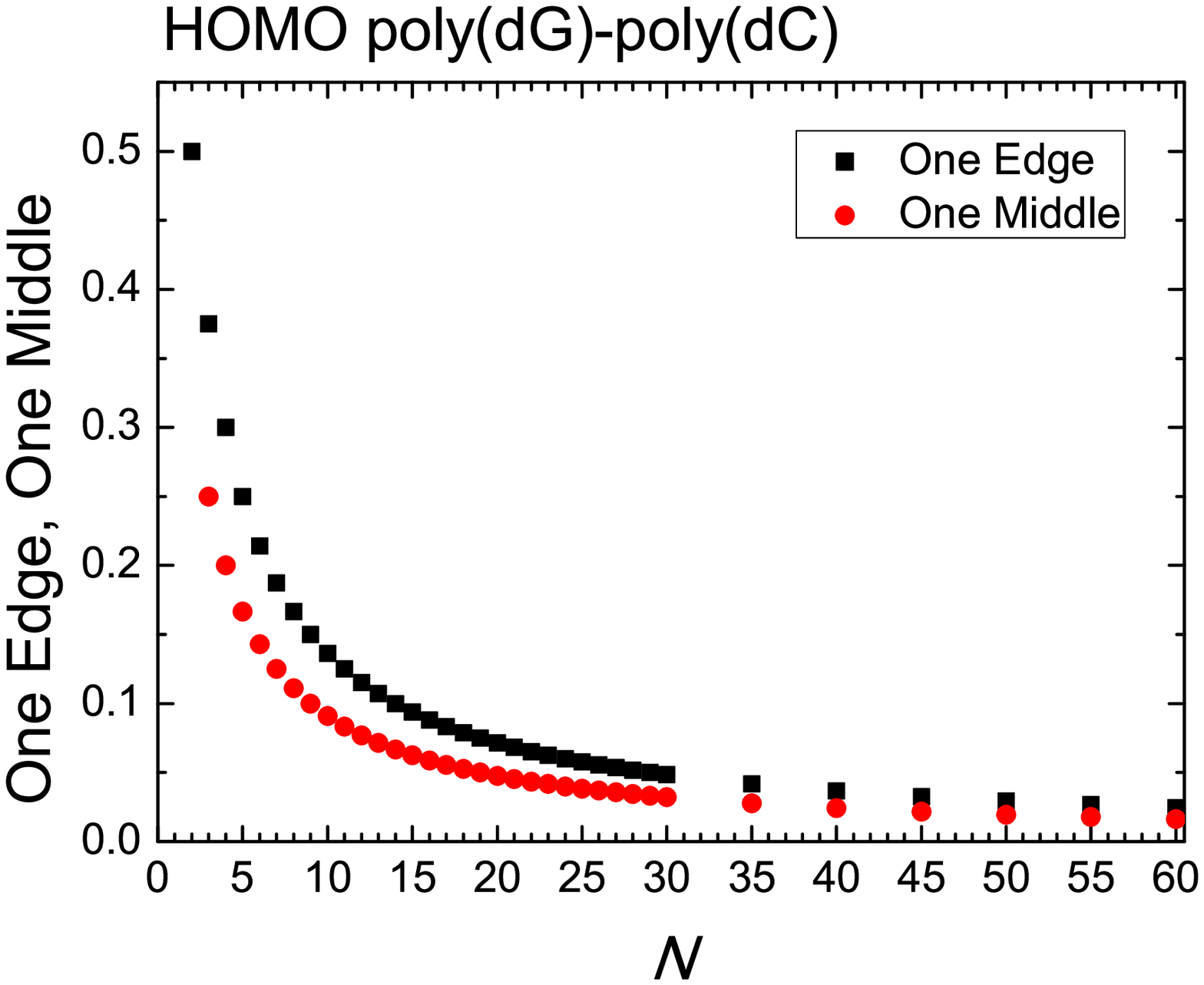}
\includegraphics[width=8cm]{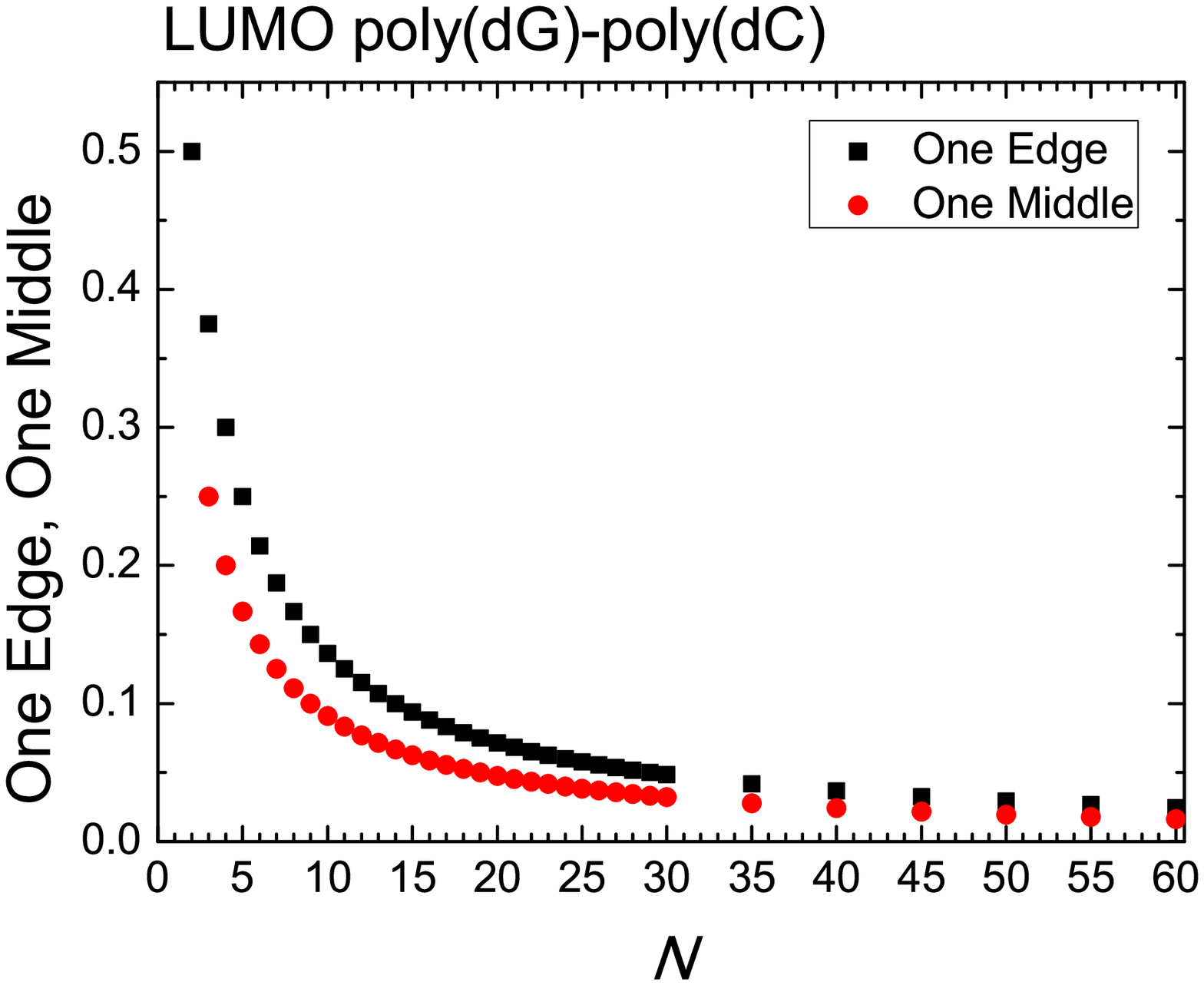}
\includegraphics[width=8cm]{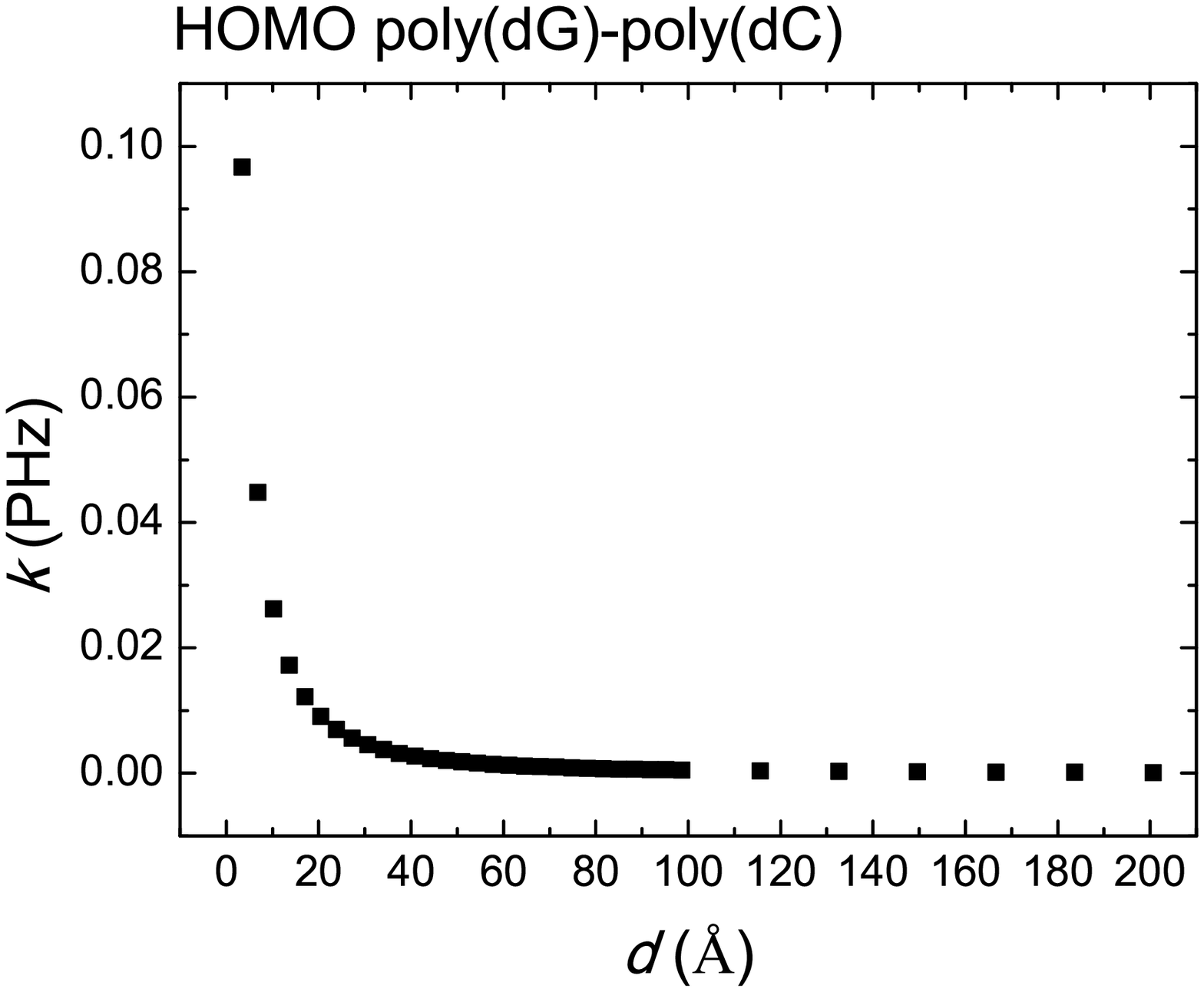}
\includegraphics[width=8cm]{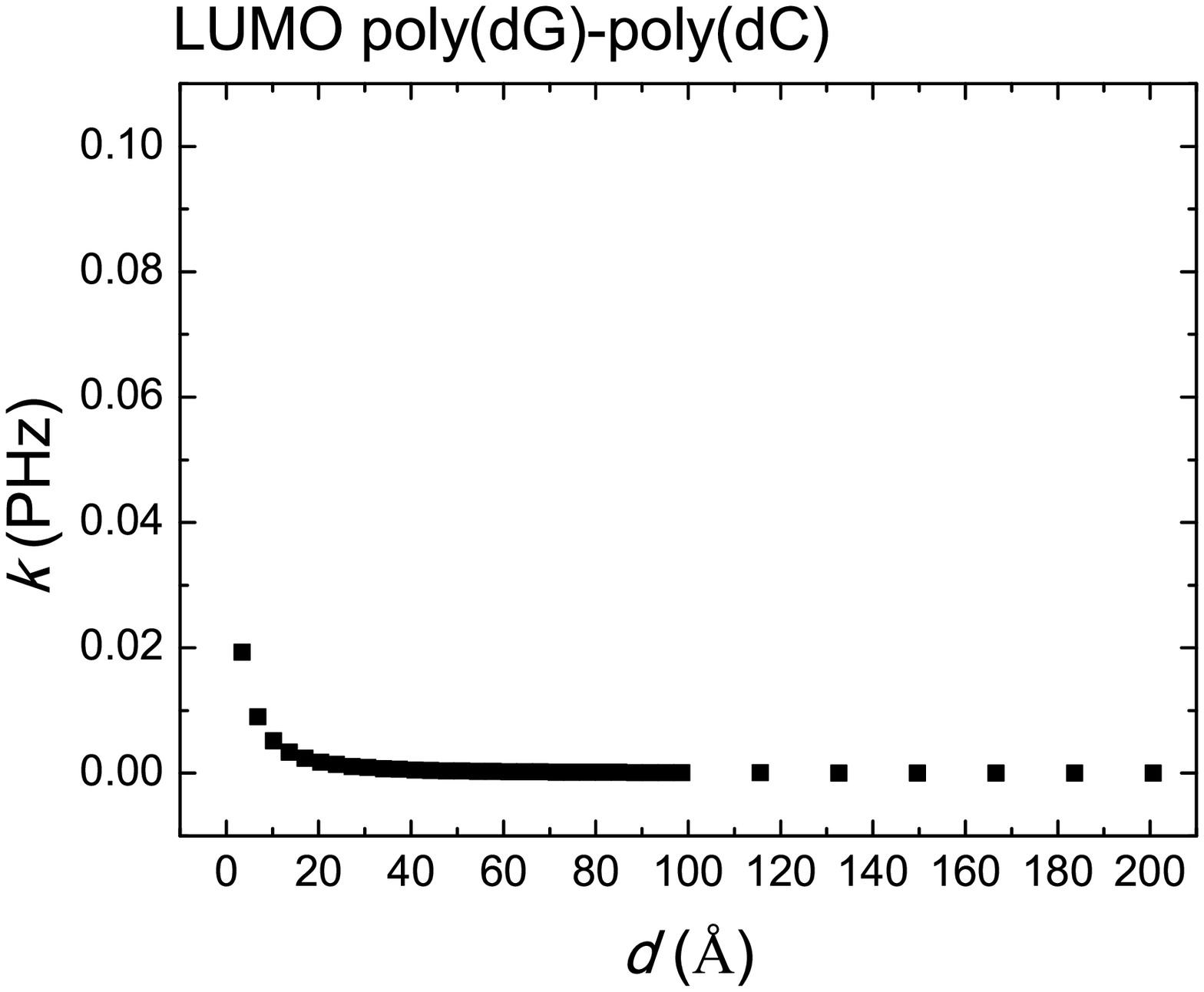}
\includegraphics[width=8cm]{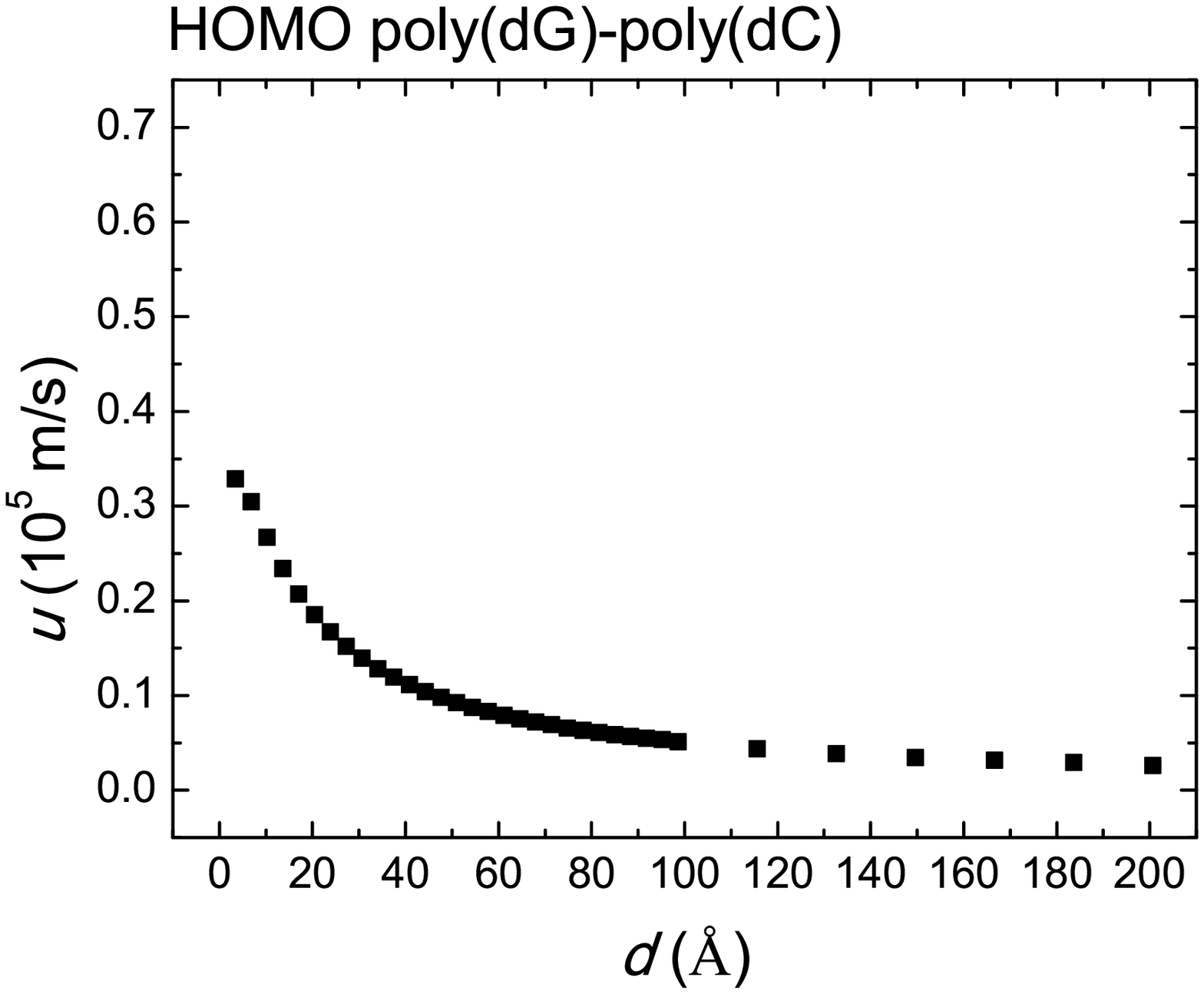}
\includegraphics[width=8cm]{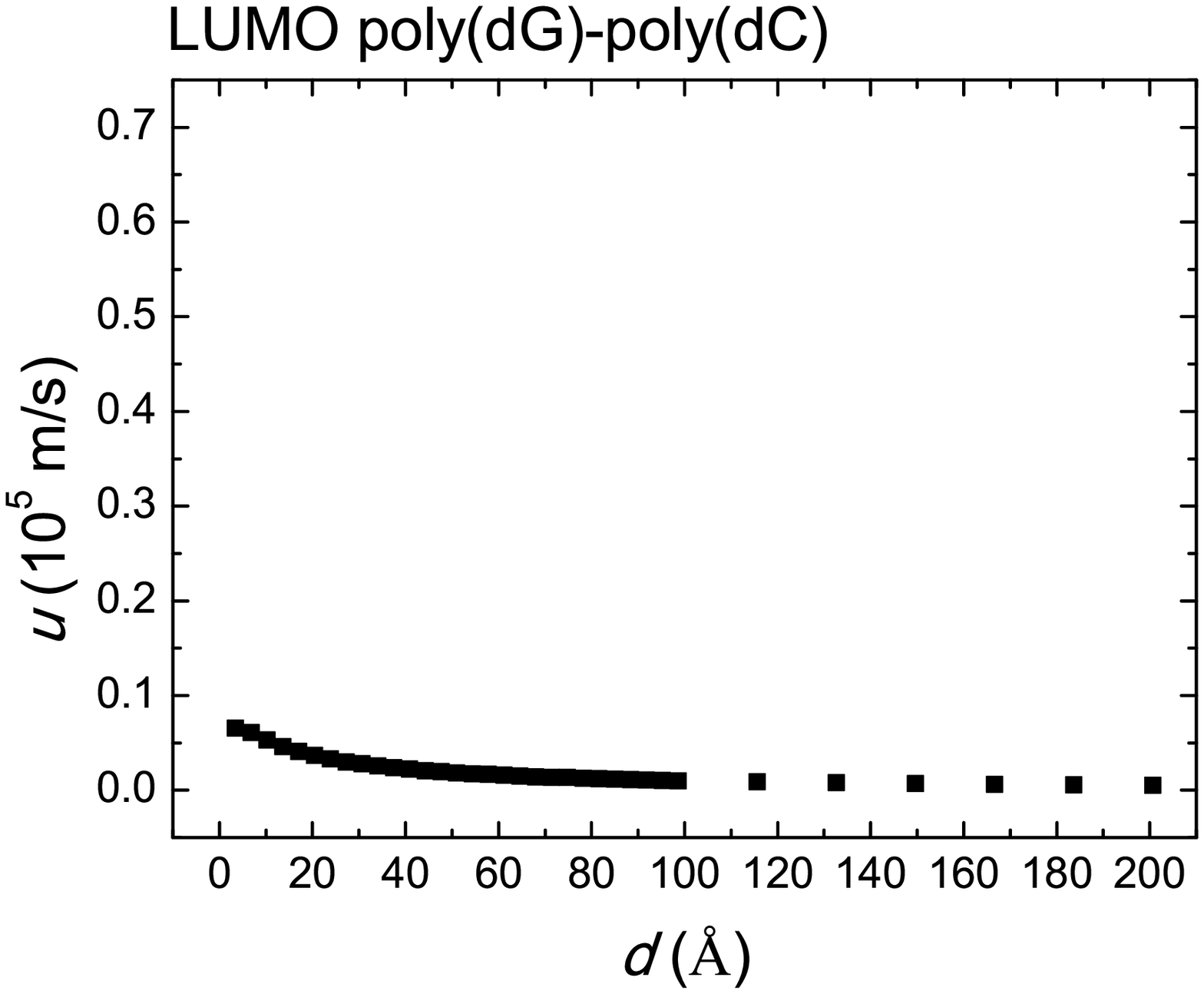}
\caption{Hole (left column) and electron (right column) transfer in \emph{poly(dG)-poly(dC)}.
[1st row] The total probability at the {\it Edge Group}, $e(N) = \frac{3}{N+1}$, and the total probability at the {\it Middle Group}, $m(N)=1-e(N)$.
[2nd row] In poly(dG)-poly(dC) these probabilities are equally distributed among the members (monomers) of these Groups.
The probability at each of the members of the Groups is shown.
[3rd row] The \textit{pure} mean transfer rate $k$ as a function of the distance from the first to the last monomer i.e. the charge transfer distance $d = N \times$ 3.4 {\AA}.
[4th row] The speed of charge transfer defined as $ u = k d$.}
\label{fig:HOMOLUMOpolydGpolydC}
\end{figure*}

\begin{figure}[]
\centering
\includegraphics[width=8cm]{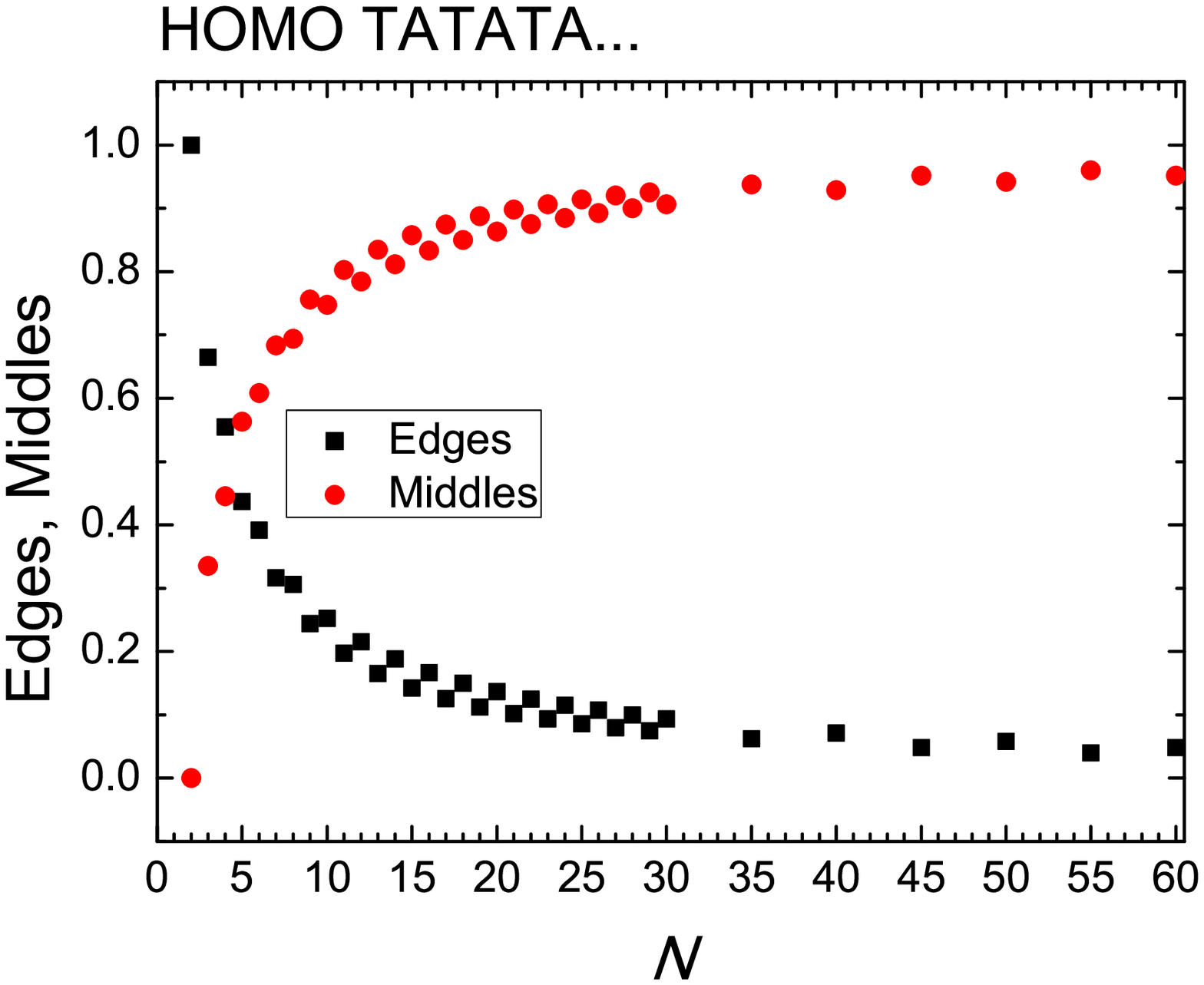}
\includegraphics[width=8cm]{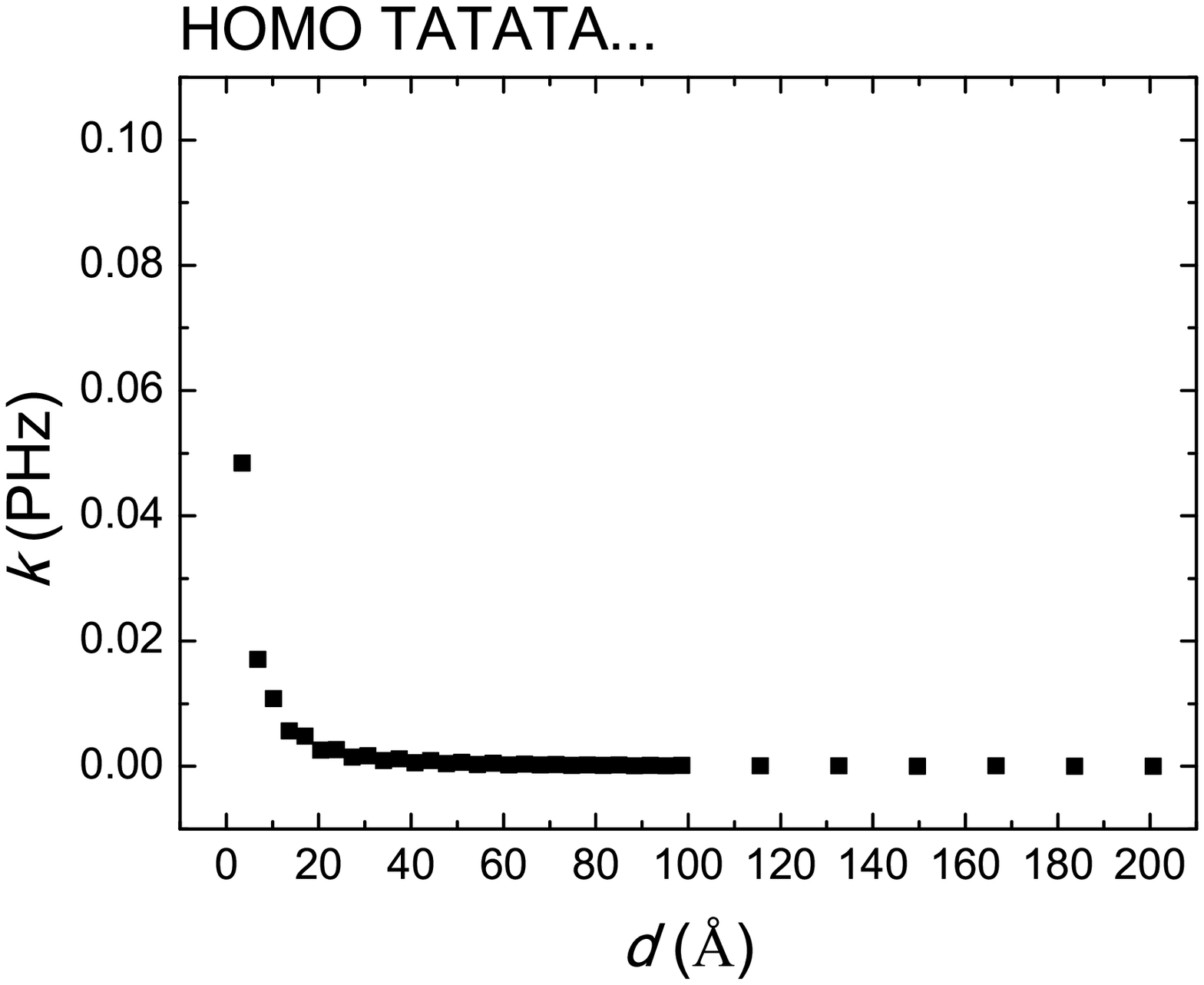}
\includegraphics[width=8cm]{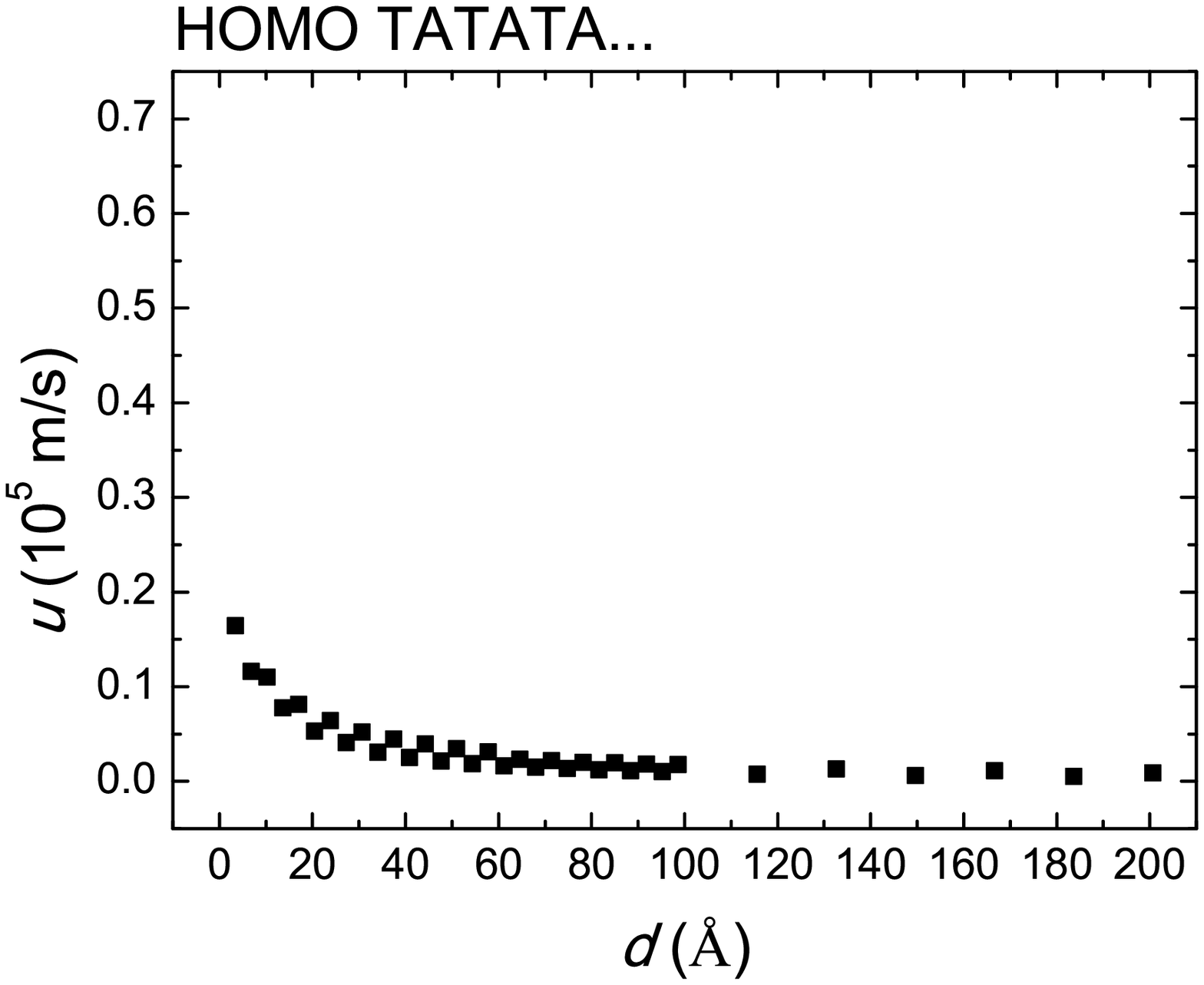}
\caption{Hole transfer in \emph{TATATA...}.
[1st panel] The total probability at the {\it Edge Group} and the total probability at the {\it Middle Group}.
[2nd panel] The \textit{pure} mean transfer rate $k$ as a function of the distance from the first to the last monomer i.e. the charge transfer distance $d = N \times$ 3.4 {\AA}.
[3rd panel] The speed of charge transfer $ u = k d$.}
\label{fig:HOMOTATATA}
\end{figure}

As an example, details of calculations for charge transfer in poly(dG)-poly(dC) are given in this paragraph.
The relevant quantities are shown  in Figure~\ref{fig:HOMOLUMOpolydGpolydC}.
Let's define
the \textit{Edge Group} of monomers made up of the first and the last monomer, and
the \textit{Middle Group} of monomers made up of the rest of the monomers.
The numerical calculations show that as a function of the total number of monomers (base-pairs), $N$,
the total probability at the Edge Group, is given by $e(N) = \frac{3}{N+1}$, and of course
the total probability at the Middle Group, $m(N)=1-e(N)$.
For poly(dG)-poly(dC) these probabilities are equally distributed among the members of the Edge Group and the Middle Group.
These quantities are displayed in the first two rows of Figure~\ref{fig:HOMOLUMOpolydGpolydC}.
Next, the \textit{pure} mean transfer rate $k$ as a function of the distance from the first to the last monomer i.e. the charge transfer distance $d = N \times$ 3.4 {\AA} is displayed in the third row. Finally, one could define the speed of charge transfer as $ u = k d$, displayed in the last row. Supposing that the \textit{pure} mean transfer rate $k$ follows an exponential dependence on the charge transfer distance $d$, and fitting as $k = k_0 \textrm{exp}(-\beta d)$, an estimation of $\beta$ and $k_0$ is obtained.
In Fig.~\ref{fig:HOMOTATATA} a few aspects of hole transfer in TATATA... are displayed.

In Reference~\cite{WangLewisSankey:2004} the authors calculated the complex band structure of poly(dA)-poly(dT) and poly(dG)-poly(dC) and obtained the energy dependence $\beta(E)$. They also found the bang gap of poly(dA)-poly(dT) (2.7 eV) and of poly(dG)-poly(dC) (2.0 eV) in accordance with previous works. Within the fundamental band gap between valence and conduction band i.e. HOMO and LUMO in our approach,
they found several $\beta(E)$ curves. Since the states with large $\beta$ values do not play a significant role in conduction they noticed that only the smallest $\beta(E)$ states, described by a semielliptical-like curve in the band-gap energy region are important. This branch reaches a maximum $\beta$ value near midgap, called the branch point, $\beta_{bp}$, approximately equal to 1.5 {\AA}$^{-1}$ for both poly(dA)-poly(dT) and poly(dG)-poly(dC). Since in molecular electronics metallic contacts are made at the two ends of the molecule and electronic current is carried by electrons tunneling from the metal with energies in the band-gap region, the branch point plays an important role in the analysis of the conductance. Although the above may hold when metal conducts are attached to the molecule, in photoinduced charge transfer experiments, we are interested in states close to the top of the valence band i.e. the HOMO or close to the bottom of the conduction band i.e. the LUMO.
As far as one can judge from the figures, for the top of the valence band of poly(dA)-poly(dT) [Fig.1a of Ref.~\cite{WangLewisSankey:2004}] $\beta \approx $ 0.4 {\AA}$^{-1}$ and for poly(dG)-poly(dC) [Fig.1b of Ref.~\cite{WangLewisSankey:2004}] $\beta \approx $ 0.2 {\AA}$^{-1}$.
These values are close to the values predicted in the present work
($\approx$ 0.2 {\AA}$^{-1}$ both for poly(dA)-poly(dT) and poly(dG)-poly(dC) cf. Table~\ref{table:ExponentialkofdFit}).
For the bottom of the conduction band of poly(dA)-poly(dT) [Fig.1a of Ref.~\cite{WangLewisSankey:2004}] $\beta \approx $ 0.5 {\AA}$^{-1}$ and for poly(dG)-poly(dC) [fig.1b of Ref.~\cite{WangLewisSankey:2004}] $\beta \approx $ 0.7 {\AA}$^{-1}$.
On the contrary, the values predicted in the present work are $\approx$ 0.2 {\AA}$^{-1}$ both for poly(dA)-poly(dT) and poly(dG)-poly(dC) cf. Table~\ref{table:ExponentialkofdFit}.
Maybe this difference is related to the fact that in Ref.~\cite{WangLewisSankey:2004}, the complex band structure is calculated
using an \textit{ab initio} tight-binding method based on density-functional theory, while,
from our point of view, we expect no difference in the $\beta$
both for poly(dA)-poly(dT) and poly(dG)-poly(dC) and both for the HOMO or the LUMO
since all these are the same type of polymers with just a different strength of interaction between the constituting monomers
(hence the $k_0$ values are different and proportional to the corresponding hopping parameters).

\begin{figure}[h!]
\hspace{0cm}
\centering
\includegraphics[width=8cm]{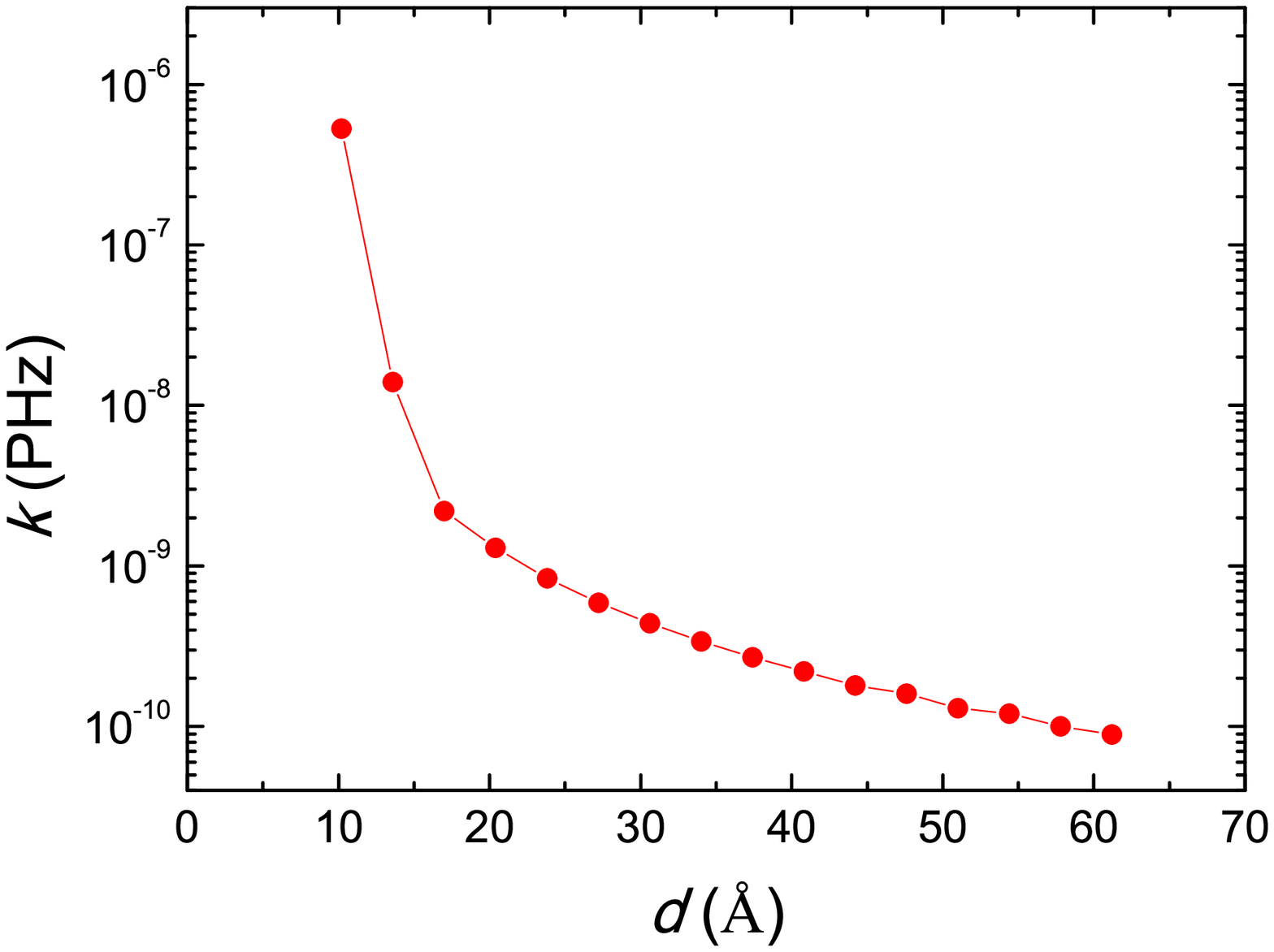}
\caption{Experiment of Giese \textit{et al.}~\cite{Giese:2001}, i.e. [G] (T)n [GGG] TATTATATTACGC. (T)n denotes the bridge made up from $n$ T-A monomers between the hole donor G (the first G-C monomer) and the hole acceptor GGG (the trimer made by three G-C monomers) before the TATTATATTACGC tail. The hole donor and acceptor are denoted by square brackets. For the strong distance dependence $k(d)$ range (for $n = $ 1,2,3), $\beta \approx$ 0.8 {\AA}$^{-1}$. In the experiment [Figure 3 of Ref.~\cite{Giese:2001}] the authors find qualitatively the same behavior, while they estimate $\beta \approx$ 0.6 {\AA}$^{-1}$ for $n = 1, 2, 3$ i.e. for the strong distance dependence range. For the weak distance dependence region, again in agreement with the experiment, a much weaker distance dependence with $\beta \approx$ 0.07 {\AA}$^{-1}$ is obtained. The line is just guide for the eyes.}
\label{fig:Giese2001kofd}
\end{figure}

In Reference~\cite{Giese:2001} Giese \textit{et al.} studied experimentally the hole transfer in the DNA segment
[G] (T)n [GGG] TATTATATTACGC. (T)n denotes the bridge made up from $n$ T-A monomers between
the hole donor G (the first G-C monomer) and the hole acceptor GGG (the trimer made by G-C monomers) before the TATTATATTACGC tail.
The hole donor and acceptor are denoted by square brackets.
In Fig.~\ref{fig:Giese2001kofd} the computed $k(d)$ i.e. the \textit{pure} mean transfer rate as a function of the distance from the hole donor to the middle of the hole acceptor is shown.
In accordance with the experiment~\cite{Giese:2001} we observe two regions with different distance dependence.
For $n = 1, 2, 3$ the distance dependence is strong becoming much weaker for $n \ge 4$.
For the strong distance dependence range, we find $\beta \approx$ 0.8 {\AA}$^{-1}$.
In the experiment [Figure 3 of Ref.~\cite{Giese:2001}] the authors find qualitatively the same behavior,
while they estimate $\beta \approx$ 0.6 {\AA}$^{-1}$ for $n = 1, 2, 3$.
For $n =4, \dots , 16$ a much weaker distance dependence is computed with $\beta \approx$ 0.07 {\AA}$^{-1}$.
Hence, the approach presented here explains the experimental results of Ref.~\cite{Giese:2001}.
%both qualitatively and quantitatively
%Moreover, we obtain almost identical results if we exclude the TATTATATTACGC tail (not shown here).
%While this manuscript was under construction Ref.~\cite{KalosakasSpanou:2013} was published where the authors, using a description of charge transfer at the single base level~\cite{HKS:2010-2011}, a different estimation of the mean transfer time involving Monte Carlo simulations [in contrast, in the present article the mean transfer time to the $j$ monomer ${t_{j}}_{mean}$ is the first time $|A_{j}(t)|^2$ becomes equal to $\langle |A_{j}(t)|^2 \rangle$], a different definition of the transfer rates being in Ref.~\cite{KalosakasSpanou:2013} just the inverse of the mean transfer time [in contrast, in the present article it is given by e.g. Eq.~\ref{meantransferrateN}, including $\langle |A_{j}(t)|^2 \rangle$], and somehow different tight-binding parameters, manage to explain qualitatively the experimental results of Ref.~\cite{Giese:2001}. However the authors notice ``a numerical discrepancy between the experimental and the calculated results is that the switching of the behavior that is obtained for $n > $ 4 in Fig. 1 is observed for $n > $ 3 in ref. 11'' and ``we see that the numerical results within our model overestimate the experimental value of $\beta$ by a factor of 2''.

In Reference~\cite{Murphy:1993} the authors demonstrated rapid photoinduced electron transfer over a distance of greater than
40 {\AA} between metallointercalators that are tethered to the 5$'$ termini of a 15-base-pair DNA duplex (5$'$-AGTGCCAAGCTTGCA-3$'$).
The authors~\cite{Murphy:1993} mentioned that ``the photoinduced electron transfer between intercalators occurs very rapidly over $>$ 40 {\AA} through the DNA helix over a pathway consisting of $\pi$-stacked base-pairs.'' Then, from Marcus theory \cite{Marcus}
they estimated $\beta$ to be $\leq$ 0.2 {\AA}$^{-1}$. We observe (Table~\ref{table:ExponentialkofdFit}) that for electron transfer (through LUMOs) we also find $\beta \leq$ 0.2 {\AA}$^{-1}$, while for hole transfer (through HOMOs) we find $\beta \approx$ 0.7 {\AA}$^{-1}$.
Similar weak distance dependence with $\beta \leq$ 0.2 {\AA}$^{-1}$ was found in Ref.~\cite{Arkin:1996}.

\begin{figure}[t!]
\hspace{0cm}
\centering
\includegraphics[width=8cm]{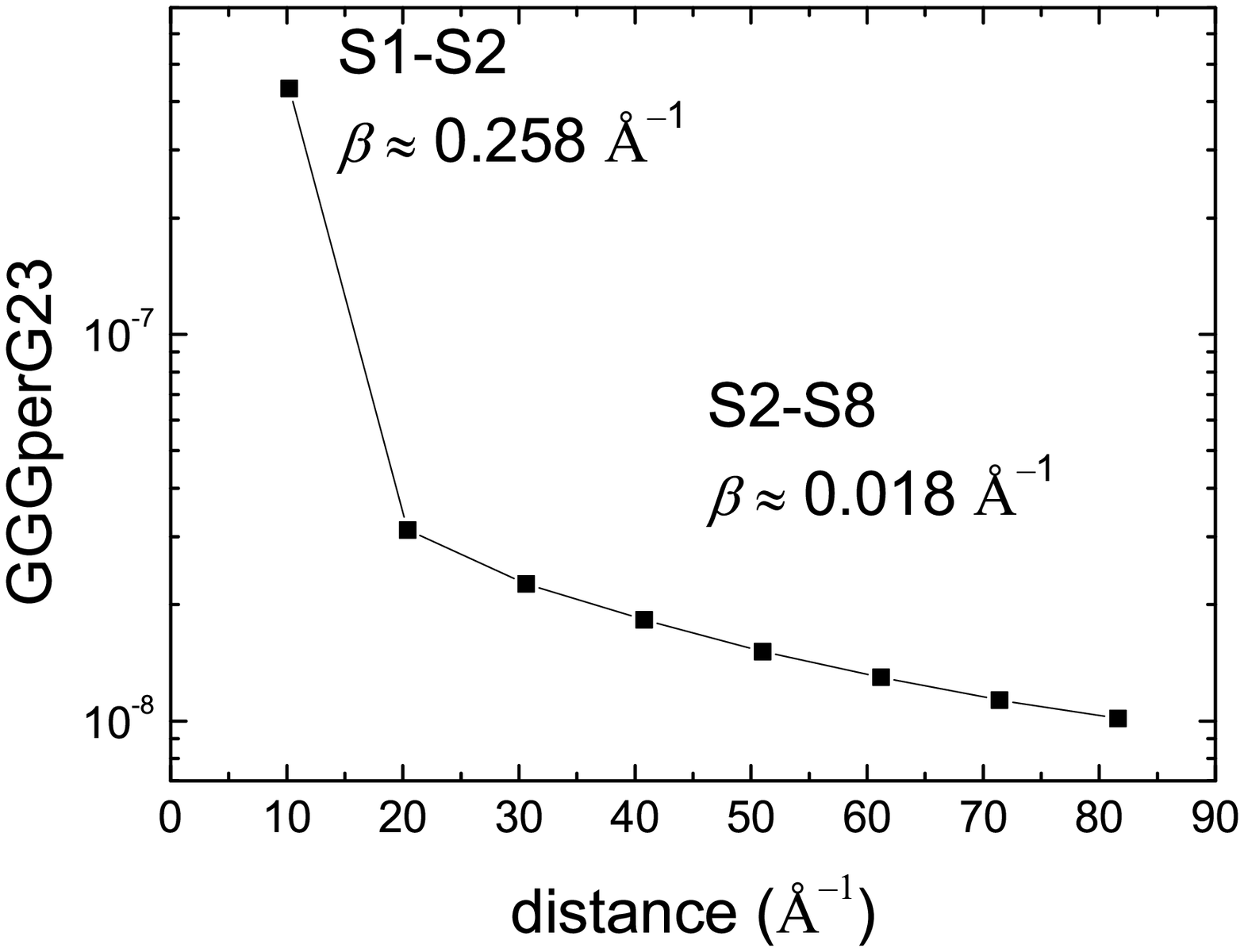}
\includegraphics[width=8cm]{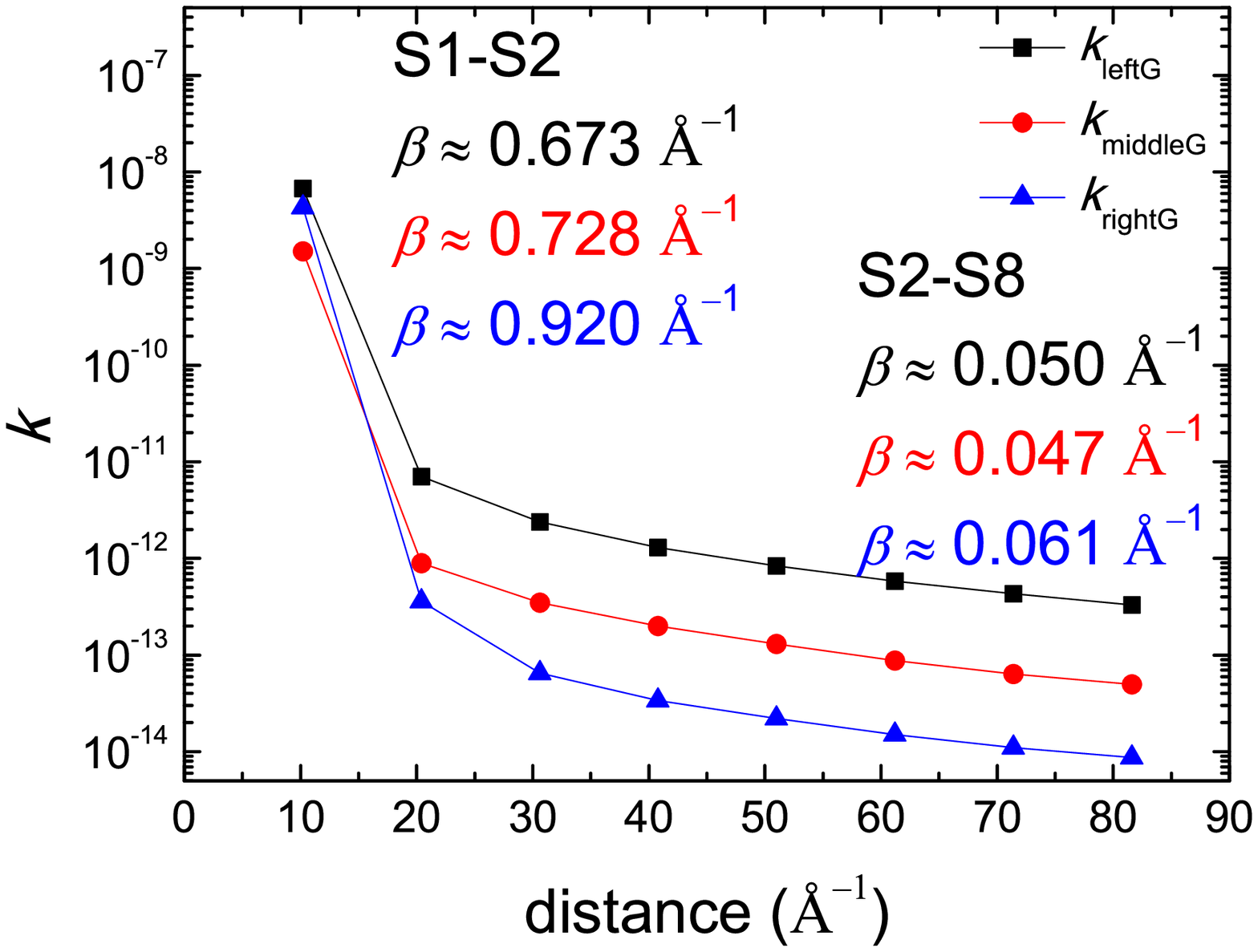}
\includegraphics[width=8cm]{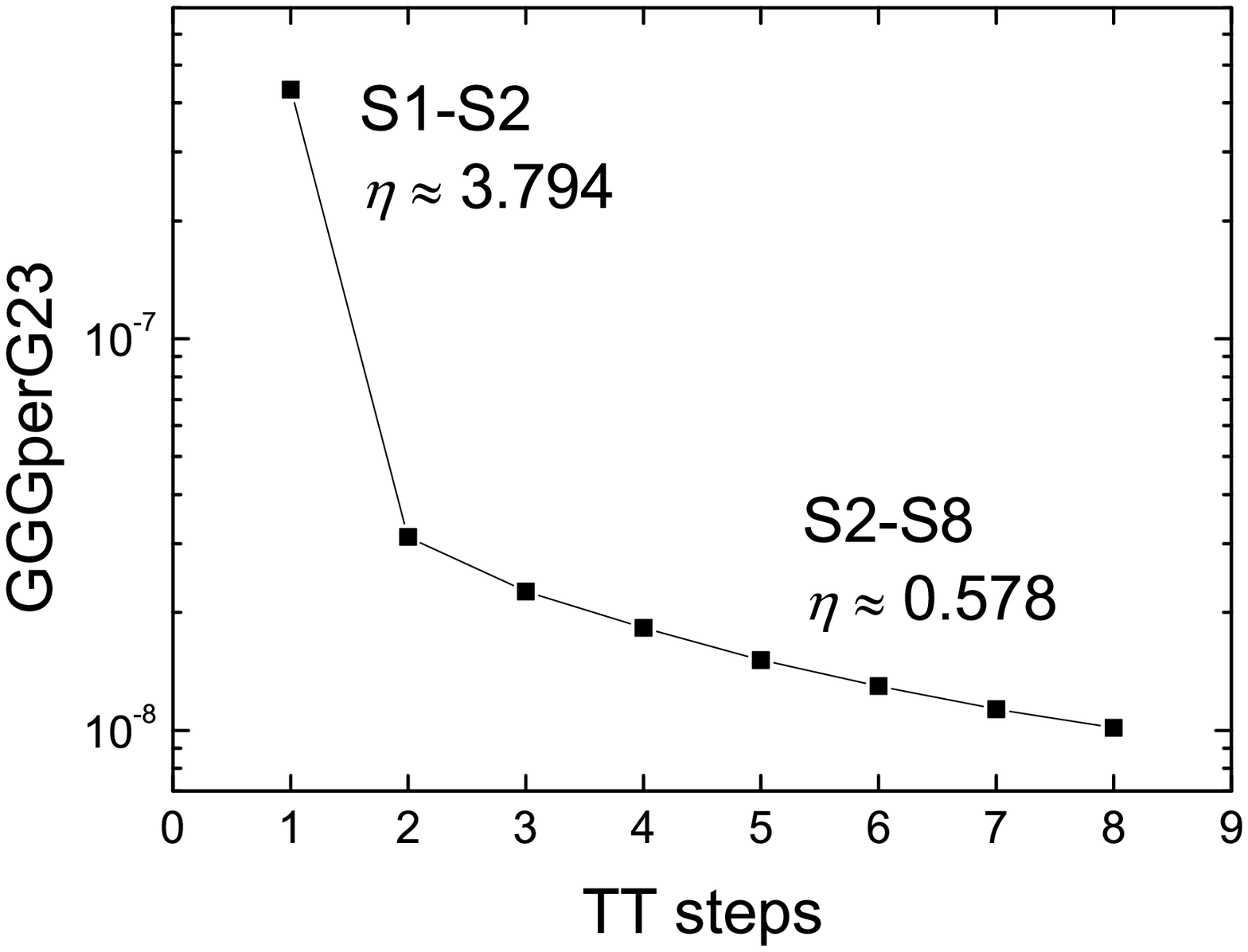}
\caption{Hole transfer in ACGCACGTCGCATAATATTACG [bridge] GGGTATTATATTACGC.
The [bridge] is made up of TT dimers separated by G monomers. In the experiment~\cite{Giese:1999}, [bridge] is either TT (one TT step) either TTGTT (two TT steps) or TTGTTGTTGTT (four TT steps). The hole is created in the C-G monomer before the G-C monomer before the [bridge], and transferred to the GGG trimer. Following Ref.~\cite{Giese:1999}, the distance is measured from G$_{23}$ to the ``left'' G of GGG. $k_{\textrm{leftG}}$, $k_{\textrm{middelG}}$, and $k_{\textrm{rightG}}$ refer to the \textit{pure} mean transfer rate of the first, second, and third G of GGG. The lines are just guides for the eyes.}
\label{fig:Giese1999}
\end{figure}
In Reference \cite{Giese:1999} the authors study hole transfer in the DNA sequence
ACGCACGTCGCATAATATTACG [bridge] GGGTATTATATTACGC,
where the [bridge] is either TT (sample 1a, one TT step) either TTGTT (sample 2a, two TT steps) or TTGTTGTTGTT (sample 3a, four TT steps).
Specifically, the hole is created in the C-G monomer before the G-C monomer before the [bridge] entity and the authors study its transfer to the GGG trimer. The charge transfer is measured by ``the oxidative damage at the G and GGG units'', ``quantified after piperidine
treatment and polyacrylamide gel electrophoresis with a phospho-imager''. Hence to compare our results with the experiment we need the ratio
of $\sum_j \langle |A_{j}(t)|^2 \rangle$ where $j$ represents the three monomers of the GGG trimer
to $\langle |A_{i}(t)|^2 \rangle$ where $i$ represents the initial G-C monomer (called also G$_{23}$).
This ratio is called GGGperG23 in Fig.~\ref{fig:Giese1999}.
% A similar comparison has been published as this manuscript was still under construction in Ref.~\cite{KalosakasSpanou:2013}, where the authors, using a description of charge transfer at the single base level~\cite{HKS:2010-2011}, a different estimation of the mean transfer time involving Monte Carlo simulations [in contrast, in the present article the mean transfer time to the $j$ monomer ${t_{j}}_{mean}$ is the first time $|A_{j}(t)|^2$ becomes equal to $\langle |A_{j}(t)|^2 \rangle$], a different definition of the transfer rates being in Ref.~\cite{KalosakasSpanou:2013} just the inverse of the mean transfer time [in contrast, in the present article it is given by e.g. Eq.~\ref{meantransferrateN}, including $\langle |A_{j}(t)|^2 \rangle$], and somehow different tight-binding parameters, manage to explain qualitatively the experimental results of Ref.~\cite{Giese:1999}.
Our calculations with three or four TT steps confirm the experimental behavior either using an exponential fit with the $\beta$ parameter or a power law fit with the $\eta$ parameter.
Extending the present approach up to eight TT steps reveals (Fig.~\ref{fig:Giese1999}) that
there are two distinct regions (i) one step (S1) to two steps (S2), and (ii) more than two steps (up to eight steps are included in the graphs).
Following Ref.~\cite{Giese:1999}, in Fig.~\ref{fig:Giese1999} the distance is measured from G$_{23}$ to the ``left'' G of GGG.
In Fig.~\ref{fig:Giese1999}, $k_{\textrm{leftG}}$, $k_{\textrm{middelG}}$, and $k_{\textrm{rightG}}$ refer to the \textit{pure} mean transfer rate of the first, second, and third G of GGG.
Finally, we underline that a description based merely on relative occupancies like the ratio GGGperG23 (cf. Fig.~\ref{fig:Giese1999} top and bottom panels), is not identical to a description based purely on  transfer rates involving the time aspect (cf. Fig.~\ref{fig:Giese1999} middle panel). Although rather obvious, this issue is obscure in the literature: often the exponential or the power fit are used either for the former or for the later case.

%%%%%%%%%%%%%%%%%%%%%%%%%%%%%%%%%%%%%%%%%%%%%%%%%%%%%%%%%%%%%%%%%%%%
\section{\label{sec:conclusion} Conclusion}
%%%%%%%%%%%%%%%%%%%%%%%%%%%%%%%%%%%%%%%%%%%%%%%%%%%%%%%%%%%%%%%%%%%%
A handy method to examine the charge transfer properties of DNA segments has been displayed. It allows to illustrate to which extent a specific DNA segment can serve as an efficient medium for charge transfer. The temporal and spatial evolution of elctrons or holes along a $N$ base-pair DNA segment can be determined, solving a system of $N$ coupled differential equations. As input one needs the relevant on-site energies of the base-pairs and the hopping parameters between successive base-pairs.
The method can be applied to any DNA segment. In the present manuscript, it has been applied to all possible dimers either for holes or for electrons, to all possible trimers for holes, and to various polymers (e.g.  poly(dG)-poly(dC), poly(dA)-poly(dT), GCGCGC..., ATATAT...) either for electrons or holes. The results have been succesfully compared with results obtained by other workers, especially experimental ones.
Some useful physical quantities were defined and calculated including the maximum transfer percentage $p$ and the \textit{pure} maximum transfer rate $\frac{p}{T}$ for cases where a period $T$ can be defined, as well as the \textit{pure} mean carrier transfer rate $k$ and the speed of charge transfer $ u = k d$, where $d = N \times$ 3.4 {\AA} is the charge transfer distance. Moreover, the inverse decay length $\beta$ used for the exponential fit $k = k_0 \textrm{exp}(-\beta d)$ and the exponent $\eta$ used for the power law fit $k = k_0' N^{-\eta}$ were computed.
The values of these parameters are not universal but depend on the specific DNA segment and are also different for electrons and holes.
Approximately, $\beta$ falls in the range $\approx$ 0.2 - 2 {\AA}$^{-1}$, $k_0$ is usually 10$^{-2}$-10$^{-1}$ PHz although, generally, it falls in the wider range 10$^{-4}$-10 PHz, while $\eta$ falls in the range $\approx$ 1.7 - 17, $k_0'$ is usually $\approx 10^{-2}$-10$^{-1}$ PHz, although generally, it falls in the wider range $\approx 10^{-4}$-10$^3$ PHz.

%\begin{acknowledgments}
%The author wishes to thank ...
%\end{acknowledgments}

\end{document}